\documentclass[aps,prb,amsfonts,amssymb,twocolumn,amsmath,preprintnumbers,nofootinbib,floatfix,
showpacs]{revtex4-1}
\usepackage[dvips]{graphics}
\usepackage{graphicx}
\usepackage{bm}
\begin{document}

\title{Influence of the spin-dependent quasiparticle distribution on the Josephson current through a ferromagnetic
weak link}
\author{A. M. Bobkov}
\affiliation{Institute of Solid State Physics, Chernogolovka,
Moscow reg., 142432 Russia}
\author{I. V. Bobkova}
\affiliation{Institute of Solid State Physics, Chernogolovka,
Moscow reg., 142432 Russia}

\date{\today}

\begin{abstract}
The Josephson current flowing through weak links containing ferromagnetic elements is studied theoretically
under the condition that the quasiparticle distribution over energy states in the interlayer is spin-dependent.
It is shown that the interplay between the spin-dependent quasiparticle
distribution and the triplet superconducting correlations induced by
the proximity effect between the superconducting leads and ferromagnetic elements of the interlayer, leads to the
appearence of an additional contribution to the Josephson current. This additional contribution $j_t$ can be extracted
from the full Josephson current in experiment. The features of the additional supercurrent $j_t$, which are
of main physical interest are the following: (i) We propose the experimental setup, where the contributions
given by the short-range (SRTC) and long-range (LRTC) components of triplet superconducting correlations
in the interlayer can be measured separately. It can be realized on the basis of S/N/F/N/S junction, where the interlayer
is composed of two normal metal regions with a spiral ferromagnet layer sandwiched between them.
For the case of tunnel junctions the measurement of $j_t$ in such a system can  provide
direct information about the energy-resolved anomalous Green's function components describing SRTC and LRTC.
(ii) In some cases the exchange field-suppressed supercurrent can be not only recovered but also enhanced
with respect to its value for non-magnetic junction with the same interface resistances by the presence of
spin-dependent quasiparticle distribution.
This effect is demonstrated for S/N/S junction with magnetic S/N interfaces. In addition,
it is also found that under the considered conditions the dependence
of the Josephson current on temperature can be nontrivial: at first the current rises upon temperature increasing
and only after that starts to decline.
\end{abstract}

\maketitle

\section{introduction}

Interplay between superconductivity and ferromagnetism in layered mesoscopic
structures offers an arena of interesting physics to explore. By now it is already
well known that so-called odd-frequency triplet pairing correlations are generated in hybrid
superconductor/ferromagnet (S/F) structures \cite{bergeret01,bergeret05}.
The essence of this pairing state is the following. The wave function of a
Cooper pair $\langle \psi_{\sigma_1}(\bm r_1,t_1)\psi_{\sigma_2}(\bm r_2, t_2)\rangle$
must be an odd function with respect to
permutations of the two electrons. Consequently, in the momentum representation the wave function of a {\it triplet}
Cooper pair has to be an odd
function of the orbital momentum for equal times $t_1=t_2$, that is, the orbital angular momentum L is an odd number.
Thus, the triplet superconducting condensate is sensitive to the presence of impurities, because only
the s-wave (L = 0) singlet condensate is not sensitive to the scattering by nonmagnetic impurities
(Anderson theorem). S/F hybrid structures are usually composed of rather impure materials.
Therefore, according to the Pauli principle equal time triplet correlations should be suppressed there.
However, another possibility
for the triplet pairing exists. In the Matsubara representation the wave function of a triplet pair
can be an odd function of the Matsubara frequency and an even function of the momentum. Then
the sum over all frequencies is zero and therefore the Pauli principle for the equal-time
wave function is not violated. These are the odd-frequency triplet pairing correlations,
which are realized in S/F structures.

If there is no a source of spin-flip processes in the considered structure (that is
the magnetizations of all the magnetic elements, which are present in
the system, are aligned with the only one axis) then the Cooper pairs penetrating
into the nonsuperconducting part of the structure consist of electrons with opposite spins.
Their wave function is the sum of a singlet component and a
triplet component with zero total spin projection $S_z=0$ on the quantization
axis. The resulting state has common origin with the famous LOFF-state \cite{larkin64,fulde64}
and can be reffered to as its mesoscopic analogue. This mesoscopic LOFF-state was predicted
theoretically \cite{buzdin82,buzdin92} and observed experimentally \cite{ryazanov01,kontos02,
blum02,guichard03,sidorenko09}.
In this state Cooper pair acquires the total momentum $2Q$ or $-2Q$ inside the ferromagnet as a response
to the energy difference between the two spin directions. Here $Q \propto h/v_F$,
where $h$ is an exchange energy and $v_F$ is
the Fermi velocity. Combination of the two possibilities
results in the spatial oscillations of the condensate wave function $\Psi (x)$ in the ferromagnet along the direction
normal to the SF interface \cite{demler97}. $\Psi_{s} (x) \propto \cos(2Qx)$
for the singlet Cooper pair and $\Psi_{t}(x) \propto \sin(2Qx)$ for the triplet Cooper pair. The same
picture is also valid in the diffusive limit. However, there is an extra decay of the condensate
wave function due to scattering in this case. In the regime $h \gg |\Delta|$, where $\Delta$ is a
superconducting order parameter in the leads, the decay length is equal to the magnetic coherence
length $\xi_F=\sqrt{D/h}$, while the oscillation period is given by
$2\pi \xi_F$. Here $D$ is the diffusion constant in the ferromagnet, $\hbar=1$ throughout the paper.
Due to the fact that the decay length $\xi_F$ is rather short (much less than the superconducting
coherence length $\xi_S=\sqrt{D/\Delta}$) the sum of $\Psi_s(x)$ and $\Psi_t(x)$ (corresponding to
$S_z=0$) can be considered as a short-range component (SRC) of the pairing correlations induced by
the proximity effect in the ferromagnet.

The situation changes if the magnetization orientation is not fixed.
The examples are domain walls, spiral ferromagnets,
spin-active interfaces, etc. In such a system not only the singlet and triplet
$S_z=0$ components exist, but also the odd-frequency triplet component with $S_z
=\pm 1$ arises in the nonsuperconducting region. The latter component penetrates
the ferromagnet over a large distance, which can be of the order of
$\xi_N=\sqrt{D/T}$ in some cases. So, this triplet component can be considered as the long-range
triplet component (LRTC). Various superconducting hybrid
structures, where LRTC can arise, were considered in the literature
(See Refs.~\onlinecite{bergeret05},~\onlinecite{golubov04},~\onlinecite{buzdin05} and references therein).
In addition, the creation of LRTC was theoretically predicted in structures
containing domain walls \cite{volkov08,fominov07}, spin-active interfaces
\cite{asano07,eschrig03}, spiral ferromagnets
\cite{volkov06,champel08,alidoust10} and multilayered SFS systems \cite{houzet07,volkov10}.
There are several experimental works, where the long-range Josephson effect
\cite{keizer06,khaire10,anwar10} and the conductance of a spiral ferromagnet attached to two
superconductors \cite{sosnin06} were measured. These results give quite convincing
evidence of LRTC existence.

Practically all the discussed above papers are devoted to investigation
of an odd-frequency triplet component under the condition that the energy distribution
of quasiparticles is equilibrium and spin-independent. However, as it was shown recently
\cite{bobkova10}, the creation of spin-dependent quasiparticle distribution in the interlayer
of SFS junction leads to appearence of the additional contribution to the Josephson
current through the junction. This additional supercurrent flows via vector part $\bm N_{j,t}$
of supercurrent-carrying density of states, which does not contribute to the Josephson
current in a junction with $s$-wave superconductor leads if the quasiparticle distribution in the
interlayer is spin-independent. Below we briefly describe how this effect arises.

The energy spectrum of the superconducting correlations is expressed in a so-called supercurrent-
carrying density of states (SCDOS) \cite{volkov95,wilheim98,yip98, heikkila02}.
This quantity represents the density of states
weighted by a factor proportional to the current that each state carries in a certain direction.
Under equilibrium conditions the supercurrent can be expressed via the SCDOS as \cite{yip98}
\begin{equation}
j \propto \int d \varepsilon N_j (\varepsilon)\tanh \varepsilon/2T
\label{supercurrent}
\enspace ,
\end{equation}
where $\varepsilon$ stands for the quasiparticle energy,
$\tanh \varepsilon/2T=\varphi(\varepsilon)$ is the equilibrium distribution function and $N_j(\varepsilon)$ is SCDOS.
In the presence of spin effects SCDOS becomes a matrix $2 \times 2$ in spin space and can be represented as
$\hat N_j=N_{j,s}+\bm N_{j,t}\bm \sigma$, where $\sigma_i$ are Pauli matrices in spin space.
Scalar in spin space part of SCDOS $N_{j,s}$ is referred to as the singlet part of SCDOS in the paper
and vector part $\bm N_{j,t}$ is referred to as the triplet part. $\bm N_{j,t}$ is directly proportional
to the triplet part of the condensate wave function. It is well known that the spin supercurrent
cannot flow through the singlet superconducting leads. Therefore, $\bm N_{j,t}$
does not contribute to the supercurrent in equilibrium. Having in mind that the triplet part of SCDOS
is even function of quasiparticle energy, one can directly see that this is indeed the case.
Otherwise, if the distribution function becomes spin-dependent, that is
$\hat \varphi(\varepsilon)=\varphi_0(\varepsilon)+\bm \varphi(\varepsilon)\bm \sigma$, the supercurrent carried by
the SCDOS triplet component $\bm N_{j,t}$ in the ferromagnet is non-zero because the scalar product
$\bm N_{j,t}(\varepsilon)\bm \varphi(\varepsilon)$ contributes to the spinless supercurrent in this case
\cite{bobkova10}.

As it is obvious from what discussed above, the spin independent nonequilibrium quasiparticle distribution
does not result in an additional contribution to the supercurrent flowing via $\bm N_{j,t}$.
However, it is worth noting here that the effect of the spin independent nonequilibrium distribution
function has been considered as well \cite{heikkila00,yip00}. It was shown that
in the limit of small exchange fields $h \ll |\Delta|$
the combined effect of the exhange field and the nonequilibrium
distribution function is also nontrivial. For instance, part of the field-suppressed supercurrent can be recovered
by adjusting a voltage between additional electrodes, which controls the distribution function.

In the present paper we continue investigation of the interplay between the triplet correlations
and spin-dependent quasiparticle distribution. As it was explained above, the simultaneous presence
of the triplet correlations and spin-dependent quasiparticle distribution in the interlayer results in
appearence of the additional contribution to the supercurrent flowing via $\bm N_{j,t}$.
In the present paper we concentrate on two features of this additional supercurrent, which are
of main physical interest and propose appropriate mesoscopic systems, where they can be observed:

(i) The additional supercurrent allows for direct measurement of the energy-resolved odd-frequency
triplet anomalous Green's function in the interlayer. The point is that for junctions with low-transparency
interfaces between the superconductor and the interlayer region $\bm N_{j,t}$ is directly proportional
to the triplet part of anomalous Green's function in the interlayer. By measuring the "non-local" conductance
(that is, the derivative of the critical current with respect to voltage $V$, which is applied
to the additional electrodes attached to
the interlayer region and controls the value of spin injection into the interlayer), one can experimentally
obtain the value of triplet part of the anomalous Green's function in the interlayer as function of energy.
As it was discussed in the introduction, the triplet correlations induced by the proximity effect in S/F
structures are odd in Matsubara frequency, that is the corresponding two-particle condensate wave function
taken at coinciding times is zero. Therefore, the direct measurement of the energy-resolved anomalous Green's
function is of great interest.

Here we propose an experimental setup, which allows for extracting from the current SRTC and LRTC contributions
and their separate observation. By measuring the "non-local" conductance one can separately obtain the
values of LRTC and SRTC triplet parts of the anomalous Green's function in the interlayer as functions of energy.
It is based on a multilayered S/N/F/N/S junction, where a layer made of a weak ferromagnetic
alloy having exchange field $\Delta << h << \varepsilon_F$ is sandwiched between two normal metal
layers. The direction of the F layer magnetization is assumed to be nonuniform in order to have a possibility
of LRTC investigation. The leads are made of dirty s-wave superconductors.

While all the experiments described in the introduction give unambigous
signatures of the fact that the odd-frequency triplet correlations do exist in hybrid SF systems, they
do not allow for direct investigation of the triplet anomalous Green function in dependence on energy.
For example, Josephson current in equilibrium is only carried by the scalar part of SCDOS $N_{j,s}$.
Surely, it is modified by the presence of the triplet component (and, in particular, manifests weakly decaying behavior
if LRTC is present in the system). However, $N_{j,s}$ is not directly proportional to the triplet
anomalous Green function, but can contain it only in a nonlinear way. The other measurable quantity in equilibrium
is the local density of states (LDOS), where the odd-frequency triplet component
manifests itself as a zero-energy peak. This effect has been
studied as for SF bilayers so as for SN bilayers with magnetic interfaces \cite{yokoyama07,asano07_2,
braude07,linder09,yokoyama09,linder10}. However, LDOS is
also not directly proportional to the triplet anomalous Green's function. The oscillating behavior
of the critical temperature as a function of an SF bilayer width (See, for example, Ref.~\onlinecite{sidorenko09}
and references therein) is also an excellent fingerprint of the triplet correlations (one-dimensional LOFF state)
presence. However, the order parameter in the singlet superconductor S is
related only to the singlet part of the anomalous Green function, which is modified by the presence
of triplet correlations, but does not allow for their direct observation.
On the other hand, in case if the quasiparticle distribution is spin-dependent, quantities,
which are directly proportional to the triplet anomalous Green function start to contribute
to experimentally observable things. Josephson current under the condition of spin-dependent
quasiparticle distribution in the interlayer is one of them.

(ii) It is well-known that ferromagnetism and singlet superconductivity are antagonistic to each other.
In overwhelming majority of situations it results in the suppression of the Josephson current through
the system with ferromagnetic elements with respect to the system with the same interface resistances but
without ferromagnetic elements. This is also valid even if LRTC is formed in the system.
In the present paper we show that in some cases the exchange field-suppressed supercurrent can be
not only recovered but also enhanced with respect to its value for non-magnetic junction with the same
interface resistances by the presence of spin-dependent quasiparticle distribution. That is, roughly speaking,
in some cases the spin-dependent quasiparticle distribution can overcompensate the suppression
of proximity-induced superconducting correlations by ferromagnetism. We demonstrate that such an effect
can be observed in S/N/S junction with magnetic interfaces.

The paper is organized as follows. In Sec.~\ref{model}
the considered model systems are described and the theoretical
framework to be used for obtaining our results is established.
in Sec.~\ref{SNFNS} we present the results of the Josephson current calculation
for a multilayered S/NFN/S system under spin-dependent quasiparticle distribution
and demonstrate how to obtain from
these data information about the structure of the odd-frequency triplet
correlations. Sec.~\ref{SNS} is devoted to consideration of SNS
junction with magnetic SN interfaces under similar conditions for the quasiparticle
distribution in the interlayer. We summarize our finding in Sec.~\ref{summary}. In Appendix \ref{A} we
represent the results for anomalous Green's function in the interlayer and all the parts of the Josephson
current for S/NFN/S junction, calculated in the framework of a particular microscopic model of N/F/N layer.
Appendix \ref{B} is devoted to a particular microscopic model of magnetic S/N interface, which we assume to
be more appropriate for the investigation of current enhancement in SNS junction.

\section{model and general scheme of calculations}

\label{model}

\begin{figure}[!tbh]
  \centerline{\includegraphics[clip=true,width=2.2in]{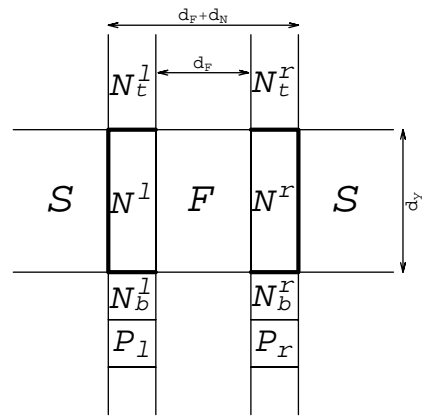}}
   \caption{S/N/F/N/S junction under consideration with the additional electrodes, which are proposed to be used
for creation of a spin-dependent quasiparticle distribution in the interlayer.}
\label{system1}
\end{figure}

The first system we consider is a multilayer S/NFN/S Josephson junction shown
schematically in Fig.~\ref{system1}. It consists of two $s$-wave superconductors (S)
and an interlayer composed of two normal layers $N^l$ and $N^r$ with a ferromagnetic
layer F, sandwiched between them. The $x$-axis is directed along the normal to the junction
and the $y$- and $z$-axes are in the junction plane. The coordinates of FN interfaces are
$x=\mp d_F/2$, while SN interfaces are located at $x=\mp(d_F+d_N)/2$. That is,  the full length
of the F layer is $d_F$, while the length of each N layer is $d_N/2$.
The middle F layer is supposed to have the exchange field $h$ satisfying the condition
$\Delta \ll h \ll \varepsilon_F$. The exchange field of F layer is assumed to be non-homogeneous,
what allows for the existence as triplet pairs with opposite spins (SRTC) so as
triplet pairs with parallel spins (LRTC) in the interlayer.
$\bm h = h (0, \sin \Theta (x), \cos \Theta (x))$, that is the magnetization vector
rotates in the F layer (within the junction plane). For simplicity we suppose that
the rotation angle has a simple $x$-dependence
\begin{equation}
\Theta (x) = \Theta' x, ~~~-d_F/2<x<d_F/2
\label {Theta_x}
\enspace ,
\end{equation}
where $\Theta'$ does not depend on coordinates. The additional electrodes are supposed to be
attached to the N layers in order to make it possible to create a spin-dependent quasiparticle distribution
in the interlayer.

We use the formalism of quasiclassical Green-Keldysh functions \cite{usadel} and assume that the superconductors
and all the internal layers are in the diffusive regime.
The fundamental quantity for diffusive transport is the momentum average of the
quasiclassical Green's function $\check g(\bm r,\varepsilon,t) =
\langle \check g(\bm p_f, \bm r,\varepsilon,t) \rangle_{\bm p_f}$. It
is a $8\times8$ matrix form in the product space of Keldysh,
particle-hole and spin variables. In the absence of an explicit dependence on time variable
the Green's function $\check g(\bm r,\varepsilon)$
in the interlayer obeys the Usadel equation
\begin{equation}
\frac{D}{\pi}\bm \nabla (\check g \bm \nabla \check g)+\left[ \varepsilon \tau_3 \sigma_0 \rho_0
- \check h , \check g \right]=0
\label{usadel}
\enspace ,
\end{equation}
where $\tau_i$, $\sigma_i$ and $\rho_i$ are Pauli matrices in particle-hole, spin and
Keldysh spaces, respectively. $\tau_0$, $\sigma_0$ and $\rho_0$ stand for the corresponding
identity matrices.
For simplicity, the diffusion constant $D$ is supposed to be identical in
all three internal layers. The matrix structure of the exchange field is as follows
\begin{equation}
\check h = \bm h \bm \sigma \rho_0 (1+\tau_3)/2 +
\bm h \bm \sigma^* \rho_0
(1-\tau_3)/2 \label{h_matrix}
\enspace .
\end{equation}
The exchange field $\bm h$ rotates in the F layer according to the described above
model. In the N layers $\bm h=0$.

The Usadel equation (\ref{usadel}) should be supplied with the normalization
condition $\check g^2 =-\pi^2 \tau_0\sigma_0\rho_0$ and is subject to Kupriyanov-Lukichev
boundary conditions \cite{kupriyanov88} at S/N and N/F interfaces.
The barrier conductances of the left and right S/N interfaces are assumed to be identical for simplicity
and are denoted as $G_T$. Then the boundary conditions at S/N interfaces take the form
\begin{equation}
\check g_{N} \partial_x \check g_{N} = -\alpha \frac{G_T}{2\sigma_N}\left[ \check g_{N}, \check g_{S} \right]
\label{boundary_SN}
\enspace .
\end{equation}
Here $\check g_N$ is the solution of the Usadel equation (\ref{usadel}) at the left ($x=-(d_N+d_F)/2$)
or right ($x=(d_N+d_F)/2$) S/N interface. $\alpha=+1(-1)$ at the left (right) interface. $\sigma_N$
is the conductivity of N layers and $\sigma_F$ is the conductivity of F layer (defined for
later use).
$\check g_S$ stands for the Green's functions at the
superconducting leads. Due to the fact that we are mostly interested in the case of low-transparent
S/N interfaces below, we can safely neglect the suppression of the
superconducting order parameter in the S leads near the interface
and take the Green's functions at the superconducting side of the
boundaries to be equilibrium and equal to their bulk values. In this case
\begin{equation}
\check g_S^K=(\check g_S^R-\check g_S^A)\tanh \frac{\varepsilon}{2T}
\label{keldysh_bulk}
\enspace ,
\end{equation}
\begin{eqnarray}
\check g_S^{R,A}=-i \pi \kappa \cosh \Theta_S^{R,A}\tau_3\sigma_0+i\pi \kappa \sinh \Theta_S^{R,A}i \sigma_2 \times
\nonumber \\
\left[e^{\displaystyle -\frac{i\alpha\chi}{2}}\frac{\tau_1+i\tau_2}{2}+
e^{\displaystyle \frac {i\alpha\chi}{2}}\frac{\tau_1-i\tau_2}{2}\right],~~~~~~~~~
\label{g_RA_bulk}
\end{eqnarray}
\begin{eqnarray}
\cosh \Theta_S^{R,A}=\frac{-\kappa
i\varepsilon}{\sqrt{|\Delta|^2-(\varepsilon+\kappa i\delta)^2}}
\nonumber \\
\sinh \Theta_S^{R,A}=\frac{-\kappa i
|\Delta|}{\sqrt{|\Delta|^2-(\varepsilon+\kappa i \delta)^2}}
\label{bulk_Theta} \enspace ,
\end{eqnarray}
where $\kappa=+1(-1)$ for the retarded (advanced) Green's function, $\chi$ stands for
the order parameter phase difference between the superconducting leads and $\delta$ is a positive infinitesimal.

The second model system, which we consider in order to study the enhancement of field-suppressed supercurrent
under the spin-dependent distribution, is an S/N/S junction with magnetic S/N interfaces.
The full length of the normal region is $d_N$, the $x$-axis is normal to the junction
plane and the interfaces are located at $x=\mp d_N/2$. As in the previous case,
additional electrodes are attached to the interlayer region for creation
of a spin-dependent quasiparticle distribution in the interlayer. The Green's function in the
N layer obeys Eq.~(\ref{usadel}) provided that $\check h = 0$. However, the boundary
conditions contain additional terms with respect to Eq.~(\ref{boundary_SN}) because
the transmission properties of spin-up and spin-down electrons into
a ferromagnetic metal or a ferromagnetic insulator are different, which gives rise
to spin-dependent conductivities (spin-filtering) and spin-dependent phase shifts
(spin-mixing) at the interface. The generalized  boundary conditions for the diffusive limit
can be written in the form \cite{huertas-hernando02,cottet09}
\begin{eqnarray}
\check g_{N} \partial_x \check g_{N} = -\alpha \frac{G_T}{2\sigma_N}\left[ \check g_{N}, \check g_{S} \right]-
~~~~~\nonumber \\
-\alpha \frac{G_{MR}}{2\sigma_N}\left[ \check g_{N}, \left\{ \check m_\alpha, \check g_{S} \right\}\right]+
\alpha \frac{G_\phi \pi}{2\sigma_N}\left[ \check m_\alpha, \check g_N \right]
\label{boundary_magnetic}
\enspace ,
\end{eqnarray}
where $\check g_N$ is the Green's function value at the normal side of the appropriate S/N interface
(at $x=\mp d_N/2$). As above, $\check g_S$ stands for the Green's function in the superconducting
lead and is expressed by Eqs.~(\ref{keldysh_bulk})-(\ref{bulk_Theta}). $\check m_\alpha=\bm m_\alpha
\bm \sigma \rho_0 (1+\tau_3)/2+\bm m_\alpha \bm \sigma^* \rho_0 (1-\tau_3)/2$, where $\bm m_\alpha$
is the unit vector aligned with the direction of the left ($\alpha=+1$) or right ($\alpha=-1$) SN
interface magnetization. $\left\{ ... \right\}$ means anticommutator. The second term accounts for
the different conductances of different spin directions and $G_{MR} \sim G_{T,\uparrow}-G_{T,\downarrow}$.
The third term $\sim G_{\phi}$ gives rise to spin-dependent phase shifts of quasiparticles being reflected at
the interface. It is worth ot note here that boundary conditions (\ref{boundary_magnetic}) are only valid
for small (with respect to unity) values of transparency and spin-dependent phase shift in one transmission
channel \cite{cottet09}. However, for the case of plane diffusive junctions we can safely consider $\tilde G_\phi
=G_\phi \xi_S/\sigma_N>1$
due to a large number of channels. $G_{\phi}$ has been calculated for some particular microscopic models of the
interface \cite{huertas-hernando02,linder10} and can be large enough even if the conductance $G_T \to 0$.
In Appendix \ref{B} we calculate $G_{\phi}$, $G_T$ and $G_{MR}$ for S/N interface
composed of an insulating barrier and a thin layer of a weak ferromagnetic alloy. We suppose this microscopic model
to be the most appropriate to the considered problem.

In what follows we assume that the S/N interfaces are low-transparent for the both considered systems,
that is $\tilde G_T \equiv G_T\xi_S/\sigma_N \ll 1$ and $\tilde G_{MR} \equiv G_{MR}\xi_S/\sigma_N \ll 1$.
In order to calculate the Josephson current through the junction in the leading order of the interface transparency
$\tilde G_T$ it is enough to obtain the retarded and advanced Green's functions in the leading order of
the transparency. If one makes use of the following definitions for the Green's function elements
in the particle-hole space (all the matrices denoted by $\hat ~$ are $2 \times 2$ matrices in spin space throughout
the paper)
\begin{equation}
\check g^{R,A}=
\left(
\begin{array}{cc}
\hat g^{R,A} & \hat f^{R,A} \\
\hat {\tilde f}^{R,A} & \hat {\tilde g}^{R,A} \\
\end{array}
\right)
\label {g_matrix}
\enspace ,
\end{equation}
then one can obtain from the Usadel equation (\ref{usadel}) and
the appropriate boundary conditions Eq.~(\ref{boundary_SN}) or
(\ref{boundary_magnetic}) that the diagonal in particle-hole space
elements of $\check g^{R,A}$ are zero-order in $\tilde G_T$ quantities
and take the following form in the interlayer
\begin{eqnarray}
\hat g^{R,A}= -i \kappa \pi
\nonumber \\
\hat {\tilde g}^{R,A}=i\kappa \pi
\label{g_zero}
\enspace .
\end{eqnarray}
The off-diagonal in particle-hole space elements of the Green's function are
of the first order in $\tilde G_T$ and should be obtained from the linearized Usadel
equations, which are to be derived from Eq.~(\ref{usadel}). It is convinient to represent the off-diagonal elements
in the following form
\begin{eqnarray}
\hat f^{R,A} = f_s^{R,A} i \sigma_2 + \bm f_t^{R,A} \bm \sigma i \sigma_2
\nonumber \\
\hat {\tilde f}^{R,A} = -i \sigma_2 \tilde f_s^{R,A} -i \sigma_2 \tilde {\bm f}_t^{R,A} \bm \sigma
\label{f_structure}
\enspace ,
\end{eqnarray}
where $f_s^{R,A}$ and $\bm f_t^{R,A}$ denote the singlet and triplet parts of the anomalous
Green's function, respectively.
For the case we consider (the magnetization vectors of all the ferromagnetic layers and spin-active interfaces,
which are present in the system, are in the junction plane) the out-of-plane $x$-component of the
triplet part is absent and the linearized Usadel equations
for the anomalous Green's function $\left\{f_s^{R,A}, \bm f_t^{R,A} \right\}$ can be written as follows
\begin{eqnarray}
2 \varepsilon f_s^{R,A} - 2 \bm h \bm f_t^{R,A} - i \kappa D \partial_x^2 f_s^{R,A}=0 ~~~
\nonumber \\
2 \varepsilon \bm f_t^{R,A} - 2 \bm h  f_s^{R,A} - i \kappa D \partial_x^2 \bm f_t^{R,A}=0
\label{usadel_lin}
\enspace .
\end{eqnarray}
According to the general symmetry relation \cite{serene83} $\hat {\tilde f}^{R,A}(\varepsilon)
={\hat f}^{R,A*}(-\varepsilon)$ the singlet and triplet parts of $\hat {\tilde f}^{R,A}$
can be expressed via the corresponding parts of $\hat f^{R,A}$ as follows
\begin{eqnarray}
\tilde f_s (\varepsilon) = -f_s^* (-\varepsilon)
\nonumber \\
\tilde {\bm f}_t (\varepsilon) = \bm f_t^* (-\varepsilon)
\label{tilde_f}
\enspace .
\end{eqnarray}

The linearized Usadel equations (\ref{usadel_lin}) should be supplemented by the appropriate
boundary condition, which are to be obtained by linearization of Eq.~(\ref{boundary_SN})
or (\ref{boundary_magnetic}) and at the S/N interfaces take the form
\begin{eqnarray}
\partial_x f_{N,s}^{R,A} = -\alpha \frac{G_T}{\sigma_N} i\kappa \pi \sinh \Theta_S^{R,A}
e^{-i\alpha \chi/2}+\alpha \frac{G_\phi}{\sigma_N}i\kappa \bm m_\alpha \bm f_{N,t}^{R,A}
\nonumber \\
\partial_x \bm f_{N,t}^{R,A} = \alpha \frac{G_\phi}{\sigma_N}i\kappa \bm m_\alpha f_{N,s}^{R,A},~~~~~~~~~~~~~~
\label{boundary_lin}
\end{eqnarray}
where $f_{N,s}^{R,A}$ and $\bm f_{N,t}^{R,A}$ are the singlet and triplet part values of
the anomalous Green's function at the normal side of the S/N interface. $G_\phi \neq 0$ only
if the S/N interface is spin-active. It is worth to note here that in this linear in $\tilde G_T$
and $\tilde G_{MR}$ approximation the term proportional to $G_{MR}$ does not enter the boundary
conditions.

Eqs.~(\ref{usadel_lin}) and (\ref{boundary_lin}) allow for the calculation
of the retarded and advanced Green's functions in the leading in transparency approximation.
However, it is not enough for obtaining of the electric current through the junction,
which should be calculated via Keldysh part of the
quasiclassical Green's function. For the plane diffusive junction
the corresponding expression for the current density reads as follows
\begin{equation}
j = \frac{-\sigma_N}{e} \int
\limits_{-\infty}^{+\infty} \frac{d \varepsilon}{8 \pi^2} {\rm Tr}_4
\left[\frac {\tau_0 + \tau_3}{2}
\left(\check g(x, \varepsilon)\partial _x \check
g(x, \varepsilon)\right)^K \right] \label{tok}
,
\end{equation}
where $e$ is the electron charge. The expression is written for the normal layer, but it is
also valid for the ferromagnetic region with the substitution $\sigma_F$ for $\sigma_N$.
$\left(\check g(x, \varepsilon)\partial_x \check
g(x, \varepsilon)\right)^K$ is  $4\times4$
Keldysh part of the corresponding combination of full Green's
function. It is convenient to calculate the current at the S/N interfaces.
Then the required combination of the Green's functions can be easily
found from Keldysh part of boundary conditions (\ref{boundary_SN})
or (\ref{boundary_magnetic}). In addition, we express Keldysh part of the full Green's function via
the retarded and advanced components
and the distribution function: $\check g^K=\check g^R \check \varphi-\check \varphi \check g^A$. Here argument
$(x, \varepsilon)$ of all the functions is omitted for brevity. The distribution function is diagonal in particle-hole
space: $\check \varphi=\hat \varphi (\tau_0+\tau_3)/2+ \sigma_2 \hat {\tilde \varphi} \sigma_2
(\tau_0-\tau_3)/2$. Then to the leading (second) order in transparency
\begin{widetext}
\begin{eqnarray}
{\rm Tr}_4
\left[\frac {\tau_0 + \tau_3}{2}
\left(\check g(x, \varepsilon)\partial _x \check
g(x, \varepsilon)\right)^K \right]=\frac{\alpha G_T i \pi}{\sigma_N}\left[ \left( \sinh \Theta^R_S+
\sinh \Theta^A_S \right)\tanh \frac{\varepsilon}{2T}\left( f_{N,s}^R e^{i\alpha \chi /2}+
\tilde f_{N,s}^A e^{-i\alpha \chi /2}\right) - \right.
\nonumber \\
\left.\left(f_{N,s}^R\tilde \varphi_0^{(0)}-\varphi_0^{(0)}f_{N,s}^A\right)
\sinh \Theta_S^Ae^{i\alpha \chi /2}+ \left(\tilde f_{N,s}^R \varphi_0^{(0)}-\tilde \varphi_0^{(0)}
\tilde f_{N,s}^A\right)\sinh \Theta_S^Re^{-i\alpha \chi /2}- ~~~~~~~~~~\right.
\nonumber \\
\left.\left(\bm f_{N,t}^R \tilde {\bm \varphi}^{(0)}-\bm \varphi^{(0)}\bm f_{N,t}^A\right)
\sinh \Theta_S^Ae^{i\alpha \chi /2}+ \left(\tilde {\bm f}_{N,t}^R \bm \varphi^{(0)}-\tilde {\bm \varphi}^{(0)}
\tilde {\bm f}_{N,t}^A\right)\sinh \Theta_S^Re^{-i\alpha \chi /2}-~~~~~~~~~~\right.
\nonumber \\
\left. 2i\pi \left( \cosh \Theta^R_S+\cosh \Theta^A_S  \right)\left(\tanh \frac{\varepsilon}{2T}-
\varphi_0^{(0)+(1)}\right) \right]+\frac{2\alpha G_{MR} i \pi}{\sigma_N}\left[ 2i\pi
\left( \cosh \Theta^R_S+\cosh \Theta^A_S  \right)\bm m_\alpha \bm \varphi^{(0)+(1)}\right]
\label{current_final}
\enspace ,
\end{eqnarray}
\end{widetext}
where $f_{N,s}^{R,A}$ and $\bm f_{N,t}^{R,A}$ are taken at the normal side of the appropriate S/N boundary.
$\varphi_0$ and $\bm \varphi$ represent the scalar and vector parts of the distribution function
$\hat \varphi=\varphi_0+\bm \varphi \bm \sigma$, which is also taken at the normal side of the appropriate
S/N boundary. The superscripts $...^{(0)}$ and $...^{(0)+(1)}$ of the distribution functions mean that
the corresponding quantity is calculated up to the zero and the first orders of magnitude in the interface
conductance $\tilde G_T$, respectively.

In order to calculate the current through the junction one should substitute Eq.~(\ref{current_final}) into
Eq.~(\ref{tok}). The resulting expression for the current can be further simplified by taking into account
the general symmetry relations between the Green's function elements \cite{serene83} expressed by Eq.~(\ref{tilde_f})
and the ones given below
\begin{eqnarray}
f_s^A(\varepsilon)=f_s^R(-\varepsilon)
\nonumber \\
\bm f_t^A(\varepsilon)=-\bm f_t^R(-\varepsilon)
\nonumber \\
\tilde \varphi_0(\varepsilon)=-\varphi_0(-\varepsilon)
\nonumber \\
\tilde {\bm \varphi}(\varepsilon) = \bm \varphi (-\varepsilon)
\label{symmetry}
\enspace .
\end{eqnarray}

Then the expression for the current density takes the form
\begin{eqnarray}
j=\!\!\int \limits_{-\infty}^{\infty}\!\! \frac{d\varepsilon}{2\pi e}\left\{ \alpha G_T
\left( {\rm Im}\left[ f_{N,s}^R e^{i\alpha \chi/2} \right]\tanh \frac{\varepsilon}{2T}{\rm Re}
\left[ \sinh \Theta_S^R \right] \right.+ \right.
\nonumber \\
\left. {\rm Re}\left[ f_{N,s}^R e^{i\alpha \chi/2} \right]\tilde \varphi_0^{(0)}
{\rm Im}\left[ \sinh \Theta_S^R \right] + \right.~~~~~~~~~~~~~~~
\nonumber \\
\left.{\rm Re}\left[ \bm f_{N,t}^R e^{i\alpha \chi/2} \right]
\tilde {\bm \varphi}^{(0)}{\rm Im}\left[ \sinh \Theta_S^R \right]  \right.-~~~~~~~~~~~~~~
\nonumber \\
\left.\pi \cosh \Theta_S^R [\varphi_0^{(0)+(1)}(\varepsilon)+\varphi_0^{(0)+(1)}(-\varepsilon)]/2
\right)+~~~~~~~~~~~
\nonumber \\
\left.\alpha G_{MR}\pi \cosh \Theta_S^R \bm m_\alpha [\bm \varphi^{(0)+(1)}(\varepsilon)+
\bm \varphi^{(0)+(1)}(-\varepsilon)]\right\}.~~~~~~~~
\label{current_simp}
\end{eqnarray}

The additional contribution to the current, which is absent for a spin-independent
distribution function, is given by the third term. As
it was mentioned in the introduction, this term (connected to the triplet part of SCDOS)
is directly proportional to the triplet
anomalous Green's function at the interface. The fifth term also results from vector part of the distribution
function, but under the considered conditions it does not contribute to the current, as it is shown below.
It is worth to note here that, as it is seen from Eq.~(\ref{current_simp}),
the singlet part of SCDOS is only expressed via the singlet part of the anomalous Green's function.
However, it does not mean that the triplet correlations do not contribute to the current for the case
of spin-independent distribution function.
They do contribute, as it was demonstrated by a number of experiments discussed in the introduction.
The point is that for the considered case of the tunnel junction $f_s$ in general contains
long-range contributions resulting from the LRTC (if they are present in the system). It is worth to emphasize
that all the aforesaid only concerns the calculation of the current at the interface. If one would calculate
the current at an arbitrary point of the interlayer, the corresponding expression would contain $\bm f_t$ quadratically.
Surely, the current by itself does not depend on $x$-coordinate, as it is required by the current conservation.

The distribution function $\hat \varphi^{(0)+(1)}$ entering current (\ref{current_simp}) should be calculated
by making use of the kinetic equation, which is obtained from the Keldysh part of Usadel equation (\ref{usadel}).
As we only need the distribution function up to the first order in the interface conductance, all the terms
accounting for the proximity effect (which are of the second order in $\tilde G_T$) drop out and the kinetic
equation takes especially simple form (we do not take into account inelastic relaxation in the interlayer)
\begin{equation}
\bm \nabla^2 \hat \varphi -\frac{i}{D}\left[ \bm h(x) \bm \sigma, \hat \varphi \right] = 0
\label{distribution_eq}
\enspace ,
\end{equation}
where the exchange field $\bm h(x)$ is determined above for the ferromagnetic layer and vanishes
for all the normal regions.

The kinetic equation should be supplemented by the boundary conditions at the S/N interfaces
and the interfaces with additional electrodes, attached to the normal regions of the interlayer
in order to create a spin-dependent quasiparticle distribution. While the boundary conditions
at the interfaces with additional electrodes are discussed below for a particular considered system,
the boundary conditions at the S/N interfaces are obtained from the Keldysh part of Eqs.~(\ref{boundary_magnetic})
or (\ref{boundary_SN}) and to the first order in the interface conductance take the form
\begin{eqnarray}
\partial_x \hat \varphi^{(0)}=\frac{\alpha i G_\phi}{2 \sigma_N}\left[ \bm m_\alpha \bm \sigma,
\hat \varphi^{(0)} \right]
\enspace , ~~~~~~~~~~~~~~~~
\nonumber \\
\partial_x \hat \varphi^{(1)}=-\frac{\alpha G_T}{2\sigma_N}\left( \cosh \Theta_S^R + \cosh \Theta_S^A \right)
\left(\! \tanh \frac{\varepsilon}{2T}-\hat \varphi^{(0)}\! \right)-~~~~
\nonumber \\
\frac{\alpha G_{MR}}{\sigma_N}\left[ \left( \cosh \Theta_S^R + \cosh \Theta_S^A \right)
\tanh \frac{\varepsilon}{2T}\bm m_\alpha \bm \sigma -~~~~~~~
\right.
\nonumber \\
\left.\cosh \Theta_S^R \bm m_\alpha \bm \sigma \hat \varphi^{(0)}-
\cosh \Theta_S^A \hat \varphi^{(0)} \bm m_\alpha \bm \sigma \right]+~~~~~~~~~~
\nonumber \\
\frac{\alpha i G_\phi}{2 \sigma_N}\left[ \bm m_\alpha \bm \sigma,
\hat \varphi^{(1)} \right]
\label{boundary_distrib_SN}
\enspace .~~~~~~~~~~~~~~~~~~
\end{eqnarray}

For the case of multilayered N/F/N interlayer Eq.~(\ref{distribution_eq}) should be also supplemented by
boundary conditions at the N/F interfaces, which are to be obtained from Keldysh part of Kupriyanov-Lukichev
boundary conditions (\ref{boundary_SN}) and are given in Appendix \ref{A} for a particular microscopic model.

\section{S/NFN/S junction}

\label{SNFNS}

Now we consider the particular systems. This section is devoted to S/NFN/S
Josephson junction. The model assumed for the exchange field of the F layer is
already described above. The anomalous Green's function is found up to the first order in S/N
conductance $\tilde G_T$ according to Eqs.~(\ref{usadel_lin}), (\ref{boundary_lin})
and boundary conditions at F/N interface, which should be easily obtained from Eq.~(\ref{boundary_SN})
for a given conductance of this interface. We assume that
the magnetization of the F layer rotates slowly, that is $\Theta' \xi_F \ll 1$,
while $\Theta' \xi_S \sim 1$ or even larger than unity. This assumption seems to
be quite reasonable \cite{bergeret05}. Therefore, upon calculating the anomalous
Green's functions we disregard all the terms proportional to $\Theta' \sqrt{D/h}$ and higher
powers of this parameter, while keeping the terms, where $\Theta'$ enters in the
dimensionless combination $\Theta' \sqrt{D/|\varepsilon|}$.

To this accuracy the triplet part of the anomalous Green's function can be represented as
\begin{eqnarray}
\bm f_t^R = \left( 0,~f_y,~f_z \right)\enspace ,~~~~~~~~~~~~
\nonumber \\
f_y = \sin \Theta(x) f_{SR}(x)-\cos \Theta(x) f_{LR}(x) \enspace ,
\nonumber \\
f_z = \cos \Theta(x) f_{SR}(x)+\sin \Theta(x) f_{LR}(x)
\label{triplet_SNFNS}
\enspace ,
\end{eqnarray}
where the $z$-axis is aligned with the direction of the exchange field in the middle of the F layer
(at $x=0$). $f_{SR}$ ($f_{LR}$) is formed by the Cooper pairs composed of the electrons with opposite
(parallel) spins. We are interested in the values of the triplet
component at the S/N interfaces, where $\sin \Theta(x) \equiv -\alpha \sin \Theta \equiv -\alpha \sin
\left[ \Theta' d_F/2 \right]$. The particular expressions for $f_{SR}$ and $f_{LR}$ and singlet component $f_s$
depend strongly on the
conductance of F/N interface and in general are quite cumbersome. In order to give an idea of their characteristic
behavior we have calculated them for the most simple model of absolutely transparent F/N interfaces.
The corresponding expressions are given in Appendix \ref{A}.

$f_{SR}$ is rapidly decaying in the interlayer, while $f_{LR}$ is
slowly decaying. Let us consider $f_{SR}$ at the left S/N
interface (the left interface is chosen just for
definiteness). It can be rewritten in the form
\begin{equation}
f_{SR}=f_{SR}^le^{-i\chi/2}+f_{SR}^re^{i\chi/2}
\label{fSR_divide}
\enspace ,
\end{equation}
where $f_{SR}^l$ is generated
by the proximity effect at the left S/N interface itself and $f_{SR}^r$ comes from the right S/N interface.
It can be shown that for thick enough ferromagnetic layer $d_F/\xi_F \gg 1$ $f_{SR}^r/f_{SR}^l$ is proportional
to the small factor $e^{-d_F/\xi_F}$. On the contrary, if $f_{LR}$ is represented as
\begin{equation}
f_{LR}=f_{LR}^le^{-i\chi/2}+f_{LR}^re^{i\chi/2}
\label{fLR_divide}
\enspace ,
\end{equation}
$f_{LR}^r$ does not contain the small factor $e^{-d_F/\xi_F}$ in the leading
approximation and, therefore, $f_{LR}$ describes the LRTC.
As it is explicitly demonstrated in Appendix \ref{A}, the characteristic decay length of $f_{LR}$
in the F layer is $|\lambda_t|^{-1}$, where $\lambda_t=\sqrt{{\Theta'}^2-2i(\varepsilon+i\delta)/D}$.
It is much larger than $\xi_F$ for the considered case $\xi_F \ll {\Theta'}^{-1}$.

To the considered accuracy the singlet component of the anomalous Green's function also
decays at the distance $\sim \xi_F$ in the F layer, just as SRTC $f_{SR}$ does, because
it is also composed of the electron pairs with antiparallel spin directions. Indeed, if $f_s^R$ at the left boundary
is also represented as
\begin{equation}
f_s^R=f_s^l e^{-i\chi /2}+ f_s^r e^{i \chi /2}
\label{fs_divide}
\enspace ,
\end{equation}
then $f_s^r \propto $ $e^{-d_F/\xi_F}$ in the limit $d_F/\xi_F \gg 1$.

Therefore, the main contribution to the Josephson current Eq.~(\ref{current_simp}) is given by
the LRTC component $f_{LR}$ of the anomalous Green's function. However, this contribution
is nonzero only for the case of spin-dependent quasiparticle distribution. In the standard
case of thermal spin-independent quasiparticle distribution the current is determined by
the singlet component $f_s$. Consequently, it only contains the term proportional to the small factor $e^{-d_F/\xi_F}$.
At first glance, it contradicts to the well-known fact that the equilibrium Josephson current
contains the contribution generated by the LRTC, if it is present in the system \cite{bergeret01, bergeret05}.
In fact, if one calculates
the current at the S/N boundary, then $f_s$ should be modified by presence of LRTC and should
contain a slowly decaying term, which provides the appropriate contribution. It is indeed the case
for the system we consider. However, the corresponding term is proportional to $(\Theta'\xi_F)^2$ and
is disregarded in our calculation. Surely, it should be taken into account upon calculating the Josephson
current for the case of spin-independent quasiparticle distribution, because in spite of the small factor
$(\Theta'\xi_F)^2$ it can result in large enough current contribution due to the absence of the suppression
factor $e^{-d_F/\xi_F}$. At the same time we can safely disregard this term, because for the considered case
of spin-dependent quasiparticle distribution the main contribution to the Josephson current is given by
$f_{LR}$ term, which contains neither $(\Theta'\xi_F)^2$ nor the ferromagnetic suppression factor $e^{-d_F/\xi_F}$.

In order to generate a spin-dependent quasiparticle distribution in the interlayer, additional electrodes are
attached to the normal regions of the interlayer. While in the paper we propose some particular way of
such a distribution creation, it is not important how particularly it is obtained. The main point
is to have a vector part $\bm \varphi(\varepsilon)$ of the distribution function in the interlayer, generated
anyway. For example, it can be created by a spin injection into the interlayer. If this is the case,
the discussed below results qualitatively survive.

In the present paper we assume that each of the normal regions of the interlayer is
attached to two additional normal electrodes $N_b^{l(r)}$ and $N_t^{l(r)}$ (See Fig.~\ref{system1}).
In their turn, the electrodes $N_b^l$ and $N_b^r$ have insertions $P_l$ and $P_r$ made of a strongly ferromagnetic
material. Let the voltage $V_b^{l(r)}-V_t^{l(r)}=V^{l(r)}$ is applied between the electrodes
$N_b^{l(r)}$ and $N_t^{l(r)}$. Here $V_b^{l(r)}$ and $V_t^{l(r)}$ are the electric potentials of the
outer regions of the $N_b^{l(r)}$ and $N_t^{l(r)}$ electrodes with respect to the potential of the superconducting leads.
It is worth to note here that the superconductor is assumed to be closed to a loop and the voltage
between the superconducting leads is absent. The conductances of the $N^{l(r)}/N_b^{l(r)}$ and $N^{l(r)}/N_t^{l(r)}$
interfaces are denoted by $g_b^{l(r)}$ and $g_t^{l(r)}$, respectively.

Further, for definiteness we consider the left normal region of the interlayer
with the corresponding additional electrodes. We choose the quantization axis $z_l$
along the magnetization of the left ferromagnetic insertion $P_l$ and the definitions $R_{P_l\uparrow}$,
$R_{P_l\downarrow}$ stand for the $P_l$ region resistivities for spin-up and spin-down electrons.
Then under the conditions that (i) the $N^l$ layer resistance $R_N$ and the resistance of $N_b^l$ region
inclosed between $N^l$ and $P_l$ can be disregarded as compared to $1/g_t$ and $R_{P_l\downarrow}$
and (ii) $1/R_{P_l\downarrow} \ll g_t^l \ll 1/R_{P_l\uparrow}$ one can believe that the voltage
drops mainly at the $P_l$ region for spin-down electrons and at the $N^l/N_t^l$ interface for spin-up electrons.
Also, the dissipative current flowing through $N_b^l/N^l/N_t^l$ system is small and can be disregarded.
Consequently, it is obtained that the
electric potentials for spin-up and spin-down electrons
in the $N_b^l$ region inclosed between $P_l$ and $N^l$ are different and practically constant over this region.
While the spin-up electrons are at the electric potential $V_b^l$ in this region, the potential for
spin-down electrons is approximately $V_t^l$.

In order to simplify the calculations we assume that $V^l=V^r =V_b-V_t\equiv 2V$. The left and the right additional electrodes
only differ by the direction of the magnetization of the $P_l$ and $P_r$ insertions. For later use we define
the unit vectors aligned with the $P_l$ and $P_r$ magnetizations as $\bm M_l$ and $\bm M_r$, respectively.
In order to satisfy the electroneutrality condition the electric potential of the superconducting leads
should be equal to $(V_t+V_b)/2$. Then the electric potentials for spin-up and spin-down electrons
in the $N_b^l$ region inclosed between $P_l$ and $N^l$ counted from the level of the superconducting leads are
$V_\uparrow=(V_b-V_t)/2=V$ and $V_\downarrow=(V_t-V_b)/2=-V$. Due to the fact that one can disregard the voltage
drop inside this region, the distribution functions for spin-up and spin-down electrons in this region are close
to the equilibrium form (with different electrochemical potentials). For the general case (if the quantization
axis does not aligned with the $P_l$ magnetization) the distribution function becomes a matrix in spin space
and takes the form
\begin{eqnarray}
\hat \varphi_l=\varphi_0 \sigma_0 + \varphi_t \bm M_l \bm \sigma
\enspace ,~~~~~~~~~
\nonumber \\
\varphi_0=\frac{1}{2}\left[\tanh \frac{\varepsilon-eV}{2T}+\tanh \frac{\varepsilon+eV}{2T}\right]
\enspace ,
\nonumber \\
\varphi_t=\frac{1}{2}\left[\tanh \frac{\varepsilon-eV}{2T}-\tanh \frac{\varepsilon+eV}{2T}\right]
\label{distribution_normal}
\enspace .
\end{eqnarray}
The same form of the distribution function is valid for the $N_b^r$ part enclosed between $P_r$ and $N^r$
with the substitution $\bm M_r$ for $\bm M_l$.

Now we can obtain the distribution function in the $N^l$ and $N^r$ regions of the interlayer, which enters
current (\ref{current_simp}). For the considered case $g_t \ll 1$ the dissipative current flowing
through $N_b^l/N^l/N_t^l$ junction is negligible and, therefore, the $y$-dependence
of the distribution function in the $N^{l(r)}$ region can be disregarded. Then under the condition
$\sigma_F \ll \sigma_N$ the distribution functions $\hat \varphi^{(0)}$ in the $N^l$ and $N^r$ regions calculated
up to the zero order in the S/N conductance $\tilde G_T$ are spatially constant and equal to
$\hat \varphi_l$ and $\hat \varphi_r$, respectively. Indeed, the equation for the distribution function
(\ref{distribution_eq}) at $\bm h=0$ and boundary conditions at S/N interfaces (\ref{boundary_distrib_SN})
(corresponding to $G_\phi =0 $) are satisfied by this solution. Boundary conditions at F/N interfaces
(\ref{boundary_distrib_FN}) are satisfied approximately due to the smallness of the distribution function
gradient under the condition $\sigma_F \ll \sigma_N$. If one goes beyond the approximation $\sigma_F/\sigma_N \ll 1$,
the distribution function in the $N^l$ and $N^r$ regions acquires gradient terms proportional to the parameter
$\sigma_F/\sigma_N$. If the F/N interface is less transparent than it is considered
in Appendix \ref{A}, the distribution function gradient in the N layer even smaller and the condition
$\sigma_F \ll \sigma_N$ is not so necessary.

Although the distribution function in the middle F layer does not enter the current expression (\ref{current_simp}),
it is interesting to discuss here how it behaves. For simplicity we consider the limiting case $\Theta' \to 0$, when
the exchange field in the ferromagnet is practically constant and the qualitative physical picture is more clear.
According to Eq.~(\ref{distribution_eq}) and boundary conditions (\ref{boundary_distrib_FN})
the scalar part of the distribution function $\varphi_0$ is constant over the F layer and coincide with
its value in the $N^l$ and $N^r$ regions. The vector component parallel to the exchange field of the ferromagnet
is a linear function of $x$-coordinate, which matches the constant values $\varphi_t \bm M_{l(r)} \bm h_{l(r)}/h$
at the F/N interfaces. Here $\bm h_{l,r} \equiv \bm h(x=\mp d_F/2)$ are the exchange field values at the left and right
N/F interfaces. The vector component perpendicular to the exchange field of the ferromagnet
decays from the F/N interfaces into the ferromagnetic region at the characteristic length $\xi_F$
oscillating simultaneously with a period $2\pi\xi_F$, as it can be obtained from Eq.~(\ref{distribution_eq})
and boundary conditions (\ref{boundary_distrib_FN}).

Strictly speaking, the distribution function in the N layers only takes form (\ref{distribution_normal})
if one assumes no spin relaxation there. Spin relaxation processes reduce vector part $\varphi_t$
of distribution function (\ref{distribution_normal}). The reduction can be roughly estimated as
$\varphi_t^{sr}=\varphi_t/(1+\tau_{esc}/\tau_{sr})$. Here $\varphi_t^{sr}$ is the vector part of the distribution
function in the presence of spin relaxation processes, while $\varphi_t$ is defined by Eq.~(\ref{distribution_normal}).
$\tau_{esc}=\sigma_N d_y /D(g_b+g_t)$ is an effective time, which an electron spends in the N layer before escaping.
$\tau_{sr}$ is the characteristic spin relaxation time. So, spin relaxation processes do not qualitatevely
influence on the distribution fuction if $\tau_{esc}/\tau_{sr} \ll 1$. This condition seems to be not restrictive in
real materials. For example, if one assume that the N layers are made of Al in normal state, where
$\lambda_{sr}=\sqrt{D\tau_{sr}}=450\mu m$ \cite{johnson85} has been reported and the condition $\tau_{esc}/\tau_{sr} =
\sigma_N d_y / (g_b+g_t) \lambda_{sr}^2 \ll 1 $ can be valid in a wide range of the values of dimensionless parameter
$(g_b+g_t)\xi_S/\sigma_N$ characterizing the joint conductance of $N_b^{l(r)}/N^{l(r)}$ and $N_t^{l(r)}/N^{l(r)}$
interfaces.

Now we turn to the discussion of the Josephson current through the junction.
It is expressed by Eq.~(\ref{current_simp}). As for the considered case
of nonmagnetic S/N interfaces $G_{MR}=0$, the last term in this formula is absent. Due to the
fact that the scalar part $\varphi_0^{(0)}$ of the distribution function in the interlayer
[Eq.~(\ref{distribution_normal})] is an odd function of quasiparticle energy, the part
of the current generated by the term $\propto \cosh \Theta_S^R \left[ \varphi_0^{(0)}(\varepsilon)+
\varphi_0^{(0)}(-\varepsilon) \right]/2$ also vanishes. Further, in order to avoid the flowing of a quasiparticle
current through the junction we assume that $|eV|<\Delta$ and the temperature is low ($T \ll \Delta$). Under these
conditions the linear in $x$-coordinate part of $\varphi^{(1)}$ (it is this term that provides the flowing of the
quasiparticle current through the junction) is zero in each of the $N$ regions of the interlayer, as it is
dictated by boundary conditions (\ref{boundary_distrib_SN}). Therefore, $\varphi^{(1)}$ is approximately
constant in each of the $N$ layers. We comment on the values of these constants below.

The first two terms in Eq.~(\ref{current_simp}) represent the contribution of the SCDOS singlet part,
which takes place as for the case of spin-independent quasiparticle distribution, so as when this distribution
is spin-dependent. We refer to this contribution as $j_s$. The particular expressions for $j_s$ can be easily found
in the framework of a given microscopic model of NFN interlayer after substitution of the particular expressions
for the singlet part of the anomalous Green's function and the scalar part of the distribution function
[Eqs.~(\ref{distribution_normal}) and (\ref{symmetry})] into Eq.~(\ref{current_simp}).

The third term in Eq.~(\ref{current_simp}) contains the current flowing through the SCDOS triplet part and
is nonzero only for the case of spin-dependent quasiparticle distribution. This contribution is the main
result of the present section. If one substitutes
the particular expressions for the triplet part of the anomalous Green's
function [Eq.~(\ref{triplet_SNFNS}) and the vector part of the distribution function
[Eqs.~(\ref{distribution_normal}) and (\ref{symmetry})] this contribution takes the form
\begin{equation}
j_t^{l,r}=-j_{SR}\frac{\bm h_{l,r} \bm M_{l,r}}{h}+\alpha j_{LR} \frac{\left( \bm M_{l,r} \times \bm h_{l,r}\right)
\bm e_x}{h}
\enspace ,
\label{jt_SNFNS}
\end{equation}
where $\bm e_x$ is the unit vector along the $x$-direction.
The currents $j_{SR}$ and $j_{LR}$ are generated by the short-range and long-range
triplet components of the anomalous Green's function, respectively. Consequently, if the F layer
is thick, that is $\xi_F \lesssim d_F$, the current $j_{SR}$ (as well as $j_s$)
is small due to the factor $e^{-d_F/\xi_F}$,
while $j_{LR}$ is not suppressed by this factor. The particular expression for $j_s$, $j_{SR}$ and $j_{LR}$ calculated
in the framework of the considered here microscopic model
are given in Appendix \ref{A}.

It is seen from Eq.~(\ref{jt_SNFNS}) that the values of the current contribution $j_t$
at the left and right S/N interfaces can be different, that is, in general,
$j_t^{l,r} = j_t \pm j_a$. However, under the condition that the superconducting leads are closed into a loop
the currents at the left and right S/N interfaces must be equal to each other. It appears
that the distribution function in the N layers acquires additional terms $\varphi^{(1)}_{l,r}$,
which are proportional to $G_T$. Under the condition $V^l=V^r$ we obtain that $\varphi^{(1)}_l=\varphi^{(1)}_r$.
Then, according to Eq.~(\ref{current_simp}) this term results in the
current contribution, which exactly compensates $j_a$. Therefore, the Josephson
current $j_t$ flowing through the junction can be simply calculated as $j_t=(j_t^l+j_t^r)/2$.

It is obvious from Eq.~(\ref{distribution_normal}) that $j_s$ is an even function of voltage $V$
applied to the additional electrodes and $j_t$ is an odd function of this voltage. Therefore,
it is easy to extract in experiment contributions $j_s$ and $j_t$ from the full Josephson current:
$j_s(V)=(j(V)+j(-V))/2$, while $j_t(V)=(j(V)-j(-V))/2$. Further, it is seen from Eq.~(\ref{jt_SNFNS})
that by choosing the appropriate orientation of $P_l$ and $P_r$ magnetizations, one can, in principle,
measure either $j_{SR}$ or $j_{LR}$ current contributions. For this reason it makes sense to discuss
all the current contributions $j_s$, $j_{SR}$ and $j_{LR}$ separately. In the tunnel limit all of them
manifest sinusoidal dependence on the superconducting phase difference $\chi$, that is $j_{s,SR,LR}=
j^c_{s,SR,LR} \sin \chi$. Therefore, we only discuss the corresponding critical currents $j^c_s$, $j^c_{SR}$,
$j^c_{LR}$ below.

\begin{figure}[!tbh]
  \centerline{\includegraphics[clip=true,width=2.2in]{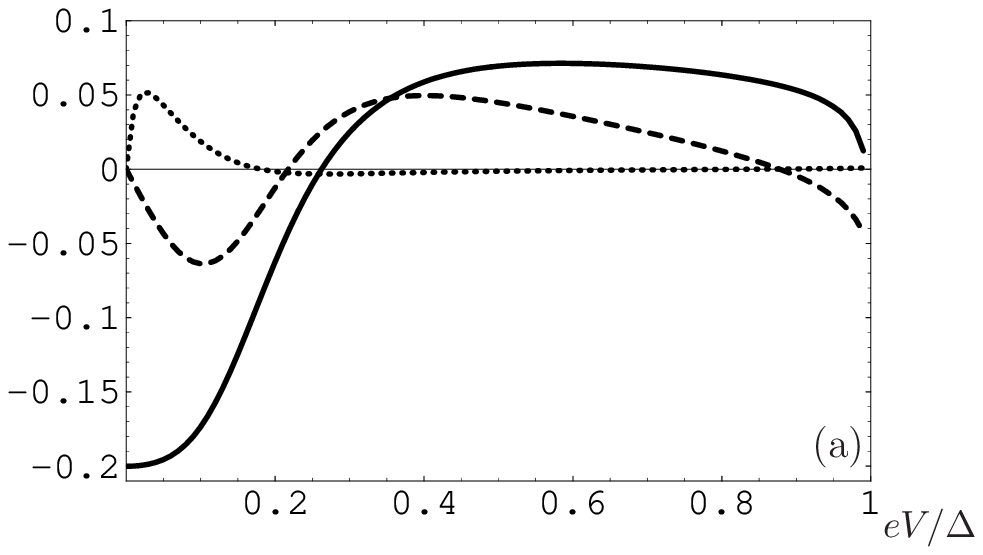}}
            \centerline{\includegraphics[clip=true,width=2.12in]{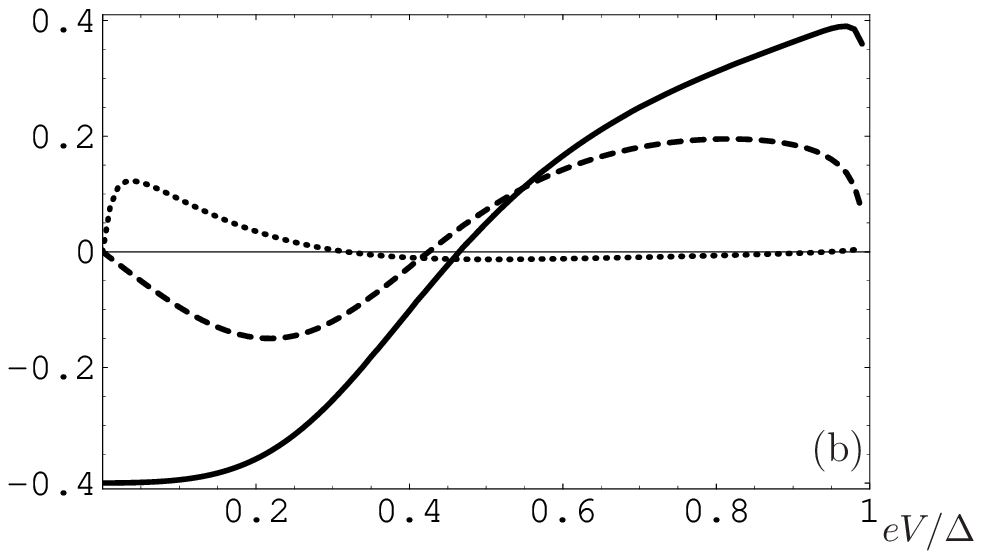}}
 \centerline{\includegraphics[clip=true,width=2.2in]{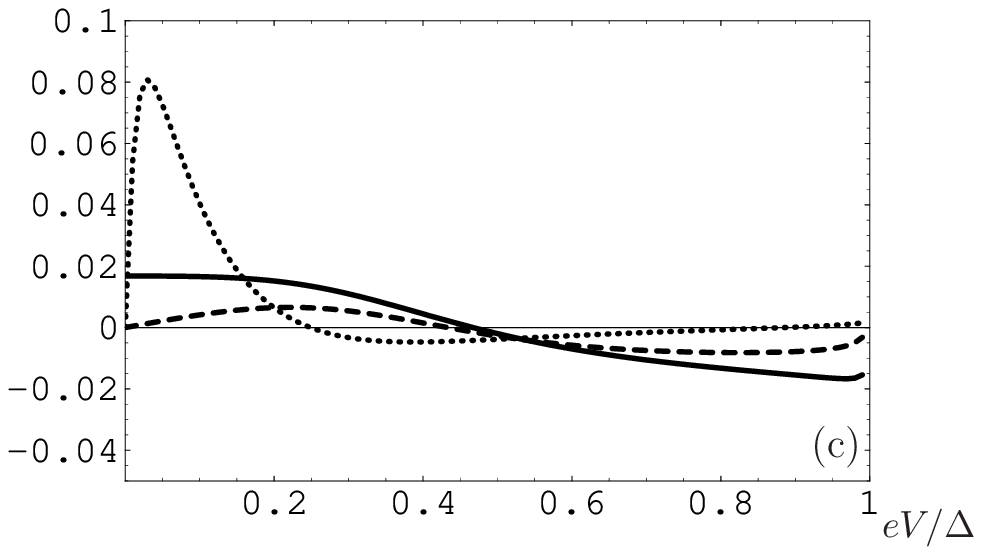}}
   \caption{Current components $j^c_s$ (solid line), $j^c_{SR}$ (dashed line) and $j^c_{LR}$ (dotted line)
as functions of voltage $V$, applied between the additional eletrodes.
The currents are measured in arbitrary units. Panel (a) corresponds to $d_F=1$, $d_N=2$, for panel (b)
$d_F=1$, $d_N=1$ and for panel (c) $d_F=2$, $d_N=1$. All lengths are measured in units of $\xi_S$. The other
parameters are the following: $h=10\Delta$, $\Theta' \xi_S=0.2$, $T=0$.}
\label{current_V}
\end{figure}

Fig.~\ref{current_V} represents these contributions, calculated in the framework of the microscopic model
discussed in Appendix \ref{A}, as a function of voltage $V$. First of all, it is worth
to note that current components $j_{SR}$ and $j_{LR}$, carried by the triplet part of SCDOS, are nonzero
only for $V \neq 0$. That is, indeed, the triplet part of SCDOS only contributes to the current if a spin-dependent
quasiparticle distribution is created in the interlayer. The exchange field $h$
is chosen to be not very strong $h=10\Delta$. Such a choice is in general agreement with the characteristic
values of the exchange field in weak ferromagnetic alloys. However, the results discussed below
qualitatively survive for the case of more strong exchange fields. Roughly speaking, increasing
of the exchange field influences the results in the same manner as increasing of the $F$ layer
length $d_F$.

Panels (a), (b) and (c) correspond to different lengths of the $N$ and $F$ regions
forming the interlayer. Below all the lengths are expressed in units of the superconducting coherence
length $\xi_S$. The magnetic coherence length $\xi_F=\xi_S \sqrt{\Delta/h}$ is approximately three times
shorter than $\xi_S$. For panels (a) and (b) the ferromagnetic layer is not thick ($d_F=1$). They
differ by the length of the normal layer: panel (a) corresponds to $d_N=2$ and for panel (b) $d_N=1$.
As it is expected, upon incresing $d_N$ the magnitude of all the current components decreases not very sharply.
The corresponding decay length is considerably larger than $\xi_F$. On the contrary,
increase of $d_F$ suppresses current components $j_s$ and
$j_{SR}$ exponentially with the characteristic decay length $\xi_F$.
It is natural because they flow via the singlet and short-range triplet components
of the anomalous Green's function. These components are composed of the Cooper pairs with opposite spin direction
and, correspondingly, decay rapidly into the depth of the ferromagnetic region. It is seen from the figures
that for panels (a) and (b) $j_{SR}$ and $j_{LR}$ are of the same order, while $j_s$ is even larger.
This is not the case for panel (c), where $d_F=2$. For this parameter range $j_s$ and $j_{SR}$ are already suppressed.
However, for a certain voltage range (small enough voltages) $j_{LR}$ is not suppressed and the dependence
of its magnitude on $d_F$ is the same as on $d_N$. For larger voltages the value of $j_{LR}$ is also
suppressed. It is interesting to note that this suppression takes place for all the panels of Fig.~\ref{current_V}
irrespective of the $F$ layer length. It is obvious that the insensitivity of $j_{LR}$ to the length of the
ferromagnetic region is a result of the fact that it is carried by Cooper pairs composed of the electrons with
parallel spins. However, the characteristic behavior of this component in dependence on $V$ (sharp maximum at
small voltages and subsequent suppression) requires an additional explanation. Such an explanation is closely
connected to the particular shape of the anomalous Green's function LRTC and is given below upon discussing
the LRTC.

\begin{figure}[!tbh]
  \centerline{\includegraphics[clip=true,width=2.2in]{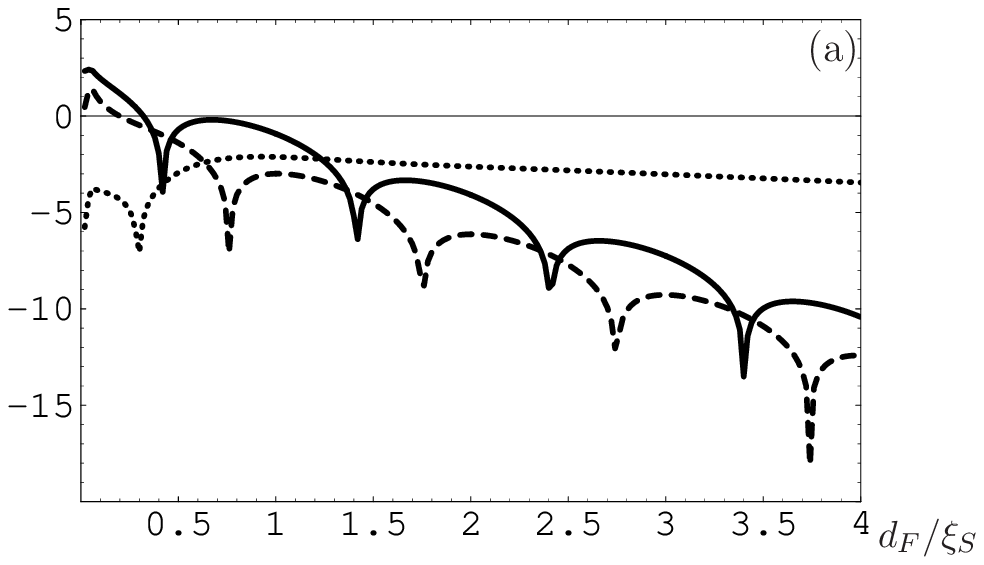}}
            \centerline{\includegraphics[clip=true,width=2.16in]{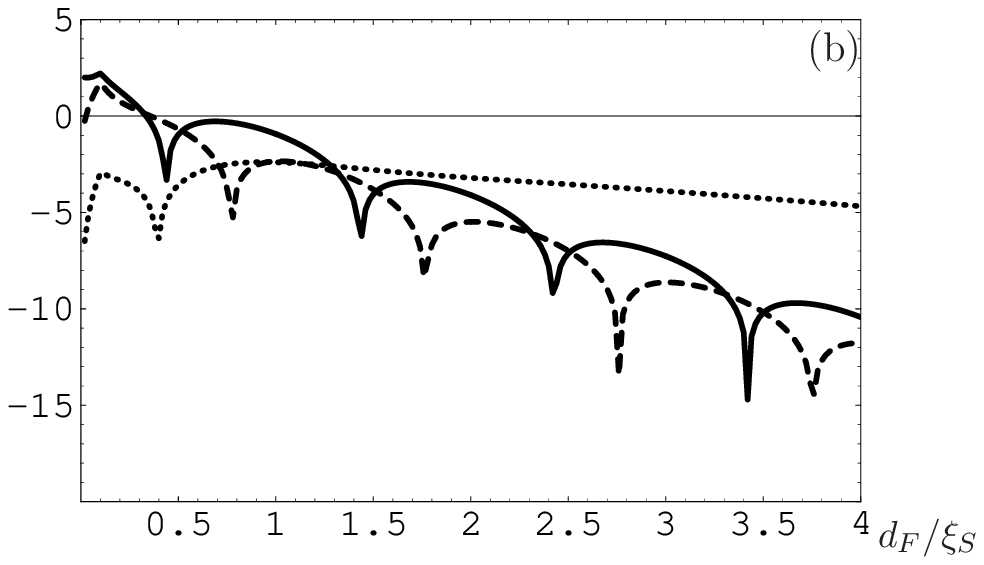}}
 \centerline{\includegraphics[clip=true,width=2.2in]{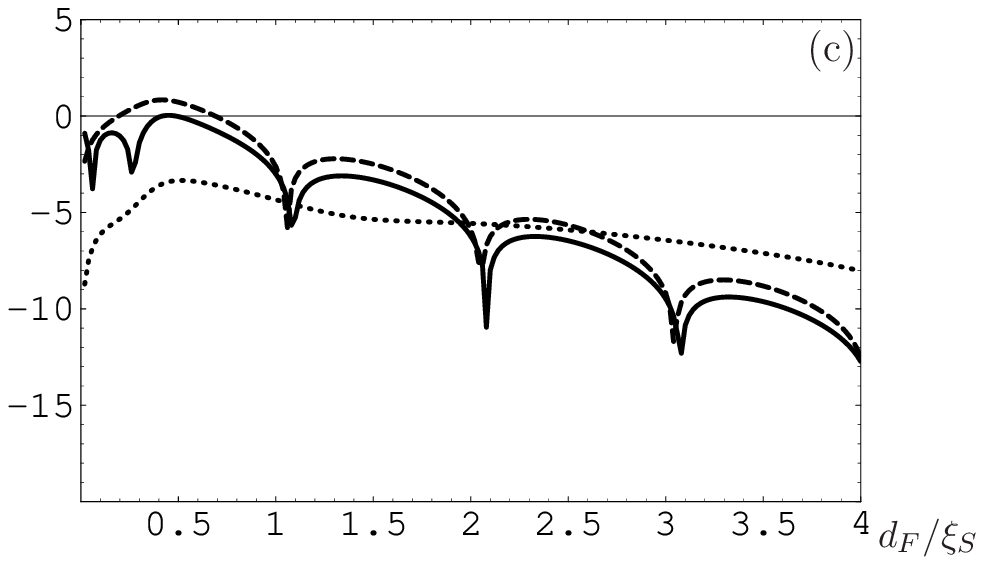}}
   \caption{Current components $j^c_s$ (solid line), $j^c_{SR}$ (dashed line) and $j^c_{LR}$ (dotted line)
in dependence on $d_F/\xi_S$ (logarithmic scale). Panel (a) corresponds to $V=0.05\Delta$, $V=0.1\Delta$ for panel (b)
and $V=0.5\Delta$ for panel (c). The other parameters are the following: $h=10\Delta$, $\Theta' \xi_S=0.2$,
$d_N=1$, $T=0$.}
\label{current_d}
\end{figure}

Further, the dependence of all three current components on the length of the ferromagnetic layer is studied
in more detail. Panels (a), (b) and (c) of Fig.~\ref{current_d} demonstrate this dependence for three different
voltages $V$. For panel (a) the particular value of this voltage is chosen to be $V=0.05\Delta$. This value
approximately corresponds to the maximum of $j_{LR}$ in Fig.~\ref{current_V}. For panel (b) $V=0.1\Delta$. Current
$j_{LR}$ gradually declines at this voltage region. Finally, the plots shown in panel (c) correspond to $V=0.5\Delta$,
where $j_{LR}$ is already greatly suppressed. First of all, it is worth to note that the decay length of $j_s$
and $j_{SR}$ is $\xi_F$ to a good accuracy for any voltage region. Also, it is seen from Fig.~\ref{current_d}
that $j_s$ and $j_{SR}$ oscillate upon increasing $d_F$ with the period $2\pi\xi_F$ (irrespective of the particular
voltage). For $j_s$, which is non-zero even for spin-independent quasiparticle distribution, these oscillations
are well-studied. They are a hallmark of mesoscopic LOFF-state, as it was mentioned in the introduction.
$j_{SR}$ is absent for spin-independent quasiparticle distribution, but is carried by the same pairs of electrons with
opposite spin directions, just as $j_s$, and, consequently, also manifests the LOFF-state oscillations.
While the oscillation period is the same for $j_s$ and $j_{SR}$, there is a phase shift between their oscillations,
which depends on the particular value of the voltage $V$.

Unlike $j_s$ and $j_{SR}$, $j_{LR}$ does not manifest oscillating behavior. Its decay length
is not connected to $\xi_F$ and crucially depends on $V$. This decay length $l_{LR}$ is maximal for the voltage region,
where $j_{LR}$ has maximal value ($l_{LR} \approx 2\xi_S \approx 6\xi_F$ for panel (a)) and declines upon increasing $V$.

\begin{figure}[!tbh]
  \centerline{\includegraphics[clip=true,width=2.5in]{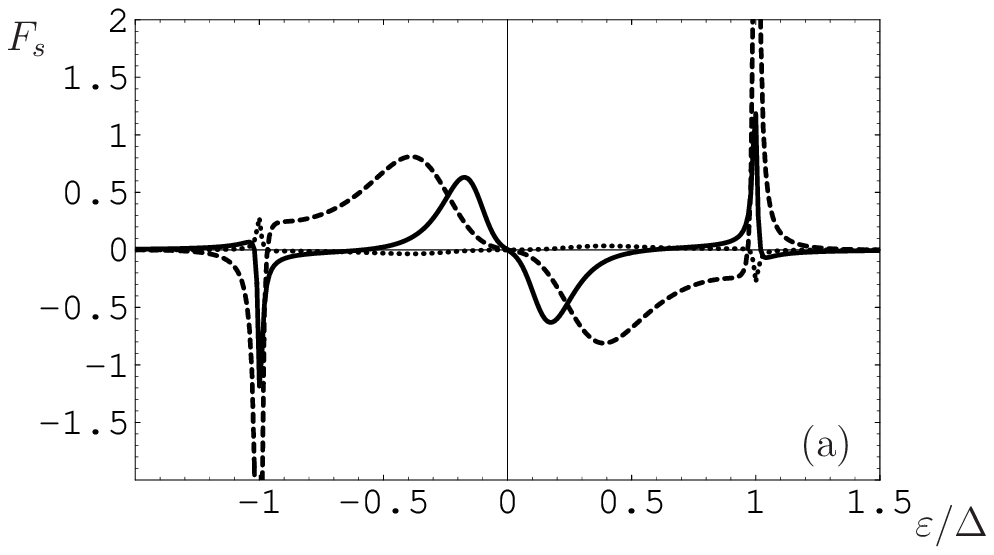}}
            \centerline{\includegraphics[clip=true,width=2.5in]{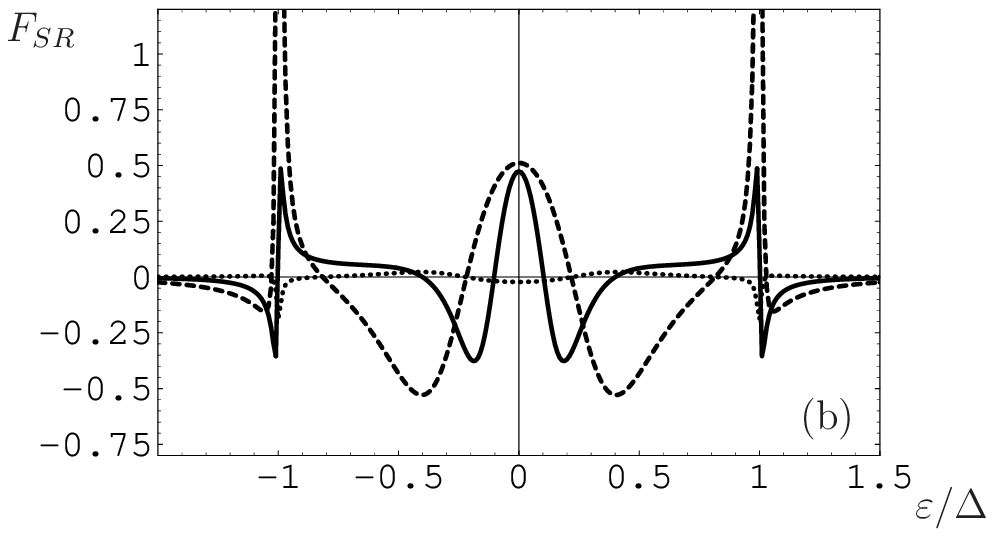}}
 \centerline{\includegraphics[clip=true,width=2.5in]{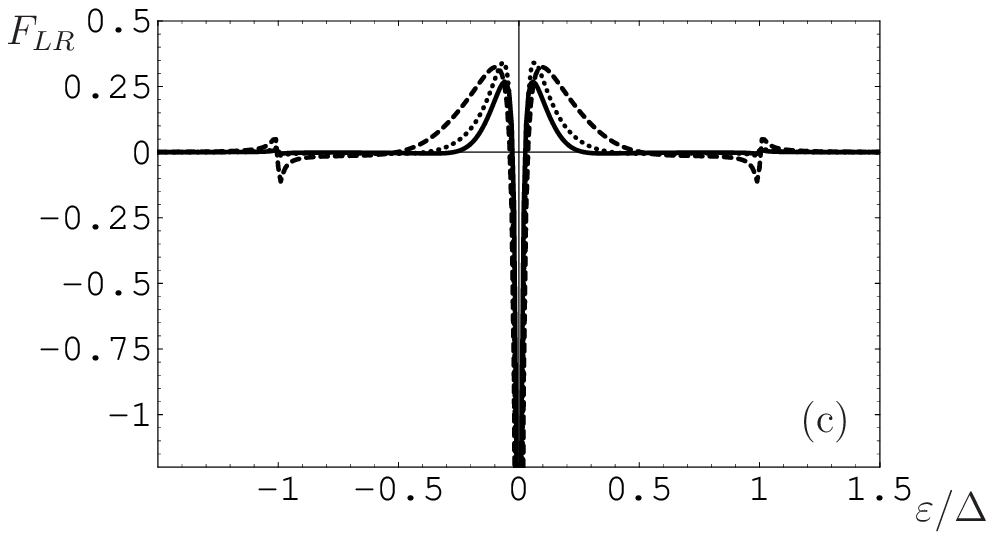}}
   \caption{Combinations $F_s$ [panel(a)], $F_{SR}$ [panel (b)] and $F_{LR}$ [panel (c)] as functions
of quasiparticle energy $\varepsilon/\Delta$. In each panel ($d_F=1$, $d_N=2$) for solid curves,
($d_F=1$, $d_N=1$) for dashed curves and ($d_F=2$, $d_N=1$) for dotted curves. The other parameters
are the same as in Fig.~\ref{current_V}.}
\label{SCDOS}
\end{figure}

As it was already mentioned in the introduction, the dependence of the anomalous Green's function in the interlayer
on the quasiparticle energy can be partially extracted from the Josephson current measurements. It can be done
due to the fact that voltage $V$ enters the current expression (\ref{current_simp}) only via distribution function
(\ref{distribution_normal}). Then, according to
Eqs.~(\ref{current_simp}), (\ref{jt_SNFNS}) and (\ref{triplet_SNFNS}), by taking the derivatives
of the Josephson currents $j_s$, $j_{SR}$ and $j_{LR}$ with respect to the voltage applied between the
additional electrodes, at $T \to 0$ one obtaines that
\begin{eqnarray}
dj_s/dV \sim
{\rm Im}\left[ \frac{f_s^r (V)}
{\sqrt{\Delta^2-(eV+i\delta)^2}}\right]
\enspace ,~~~~~~
\nonumber \\
dj_{SR}/dV \sim
{\rm Im}\left[ \frac{f_{SR}^r (V)}
{\sqrt{\Delta^2-(eV+i\delta)^2}}\right]
\enspace ,~~~~~~
\nonumber \\
dj_{LR}/dV \sim
{\rm Im}\left[ \frac{f_{LR}^r (V)}
{\sqrt{\Delta^2-(eV+i\delta)^2}}\right]
\enspace .~~~~~~~
\label{extracted}
\end{eqnarray}
Here $f_{SR}^r$, $f_{LR}^r$ and $f_s^r$ are determined by Eqs.~(\ref{fSR_divide}), (\ref{fLR_divide})
and (\ref{fs_divide}), respectively. That is, indeed, imaginary parts of the anomalous Green's function
components coming from the opposite interface corresponding to all three types of superconducting correlations
can be extracted from the Josephson current measurements.
However, under the condition $|eV|<\Delta$ it can be done only for subgap energies $|\varepsilon|<\Delta$.
It is worth to note here that the derivatives of current components $j_s$,
$j_{SR}$ and $j_{LR}$ with respect to $V$ give us the corresponding anomalous Green's function components
only in the tunnel limit $\tilde G_T \ll 1$. In general case these derivatives are proportional to the appropriate
components of SCDOS, which are expressed via the anomalous Green's function in a more complicated way.

Panels (a), (b) and (c) of Fig.~\ref{SCDOS} represent combinations $F_s \equiv
{\rm Im}\left[ f_s^r (\varepsilon)/\sqrt{\Delta^2-(\varepsilon+i\delta)^2}\right]$,
$F_{SR} \equiv {\rm Im}\left[ f_{SR}^r (\varepsilon)/
\sqrt{\Delta^2-(\varepsilon+i\delta)^2}\right]$ and $F_{LR} \equiv
{\rm Im}\left[ f_{LR}^r (\varepsilon)/\sqrt{\Delta^2-(\varepsilon+i\delta)^2}\right]$,
calculated in the framework of our microscopic model,
in dependence on quasiparticle energy $\varepsilon$ measured in units of $\Delta$. In each panel
different curves correspond to different lengths $d_N$ and $d_F$ (See caption to Fig.~\ref{SCDOS}). It is seen that
the value of the normal region length does not influence qualitatively on all three components of the
anomalous Green's function. As it is expected, $F_s$ and $F_{SR}$ are strongly suppressed upon
increasing of $d_F$. On the contrary, $F_{LR}$ is only very weakly sensitive to changing of $d_F$. It is dominated
by the sharp dip at low energies, which followed by wider peaks, where $F_{LR}$ changes sign. The width
$\delta \varepsilon$ of the dip is $\sim \sqrt{\Delta D}\Theta'$.

It is the characteristic shape of $F_{LR}$ that is responsible for $j_{LR}$ behavior in dependence on $V$, shown in
Fig.~\ref{current_V}. The point is that at low enough temperatures only the part of $F_{LR}$, belonging
to energy interval $\left[ -|eV|, |eV| \right]$, contributes to $j_{LR}$. Consequently, upon
inreasing of $V$ $j_{LR}$ grows sharply up to $V \sim (1/2)\sqrt{\Delta D}\Theta'$ and after that starts to decline
due to the opposite sign contribution of the peaks. It appears that the contributions of the dip and the peaks
mainly compensate each other, what leads to strong supression of $j_{LR}$ for large enough $V$. The discussed above
dependence of $j_{LR}$ decay length on $V$ is also closely connected
to the fact that only the part of $F_{LR}$, belonging to energy interval $\left[ -|eV|, |eV| \right]$,
"works" upon creating $j_{LR}$. Indeed, the characteristic
decay length of $F_{LR}(\varepsilon)$ $d_F^{cr}(\varepsilon) \sim |\lambda_t(\varepsilon)|^{-1}$. Therefore,
$j_{LR}$ decay length $\sim 1/\Theta'$ for small voltages and gets shorter for larger voltages due to
increased contribution of higher energies.

\begin{figure}[!tbh]
  \centerline{\includegraphics[clip=true,width=2.2in]{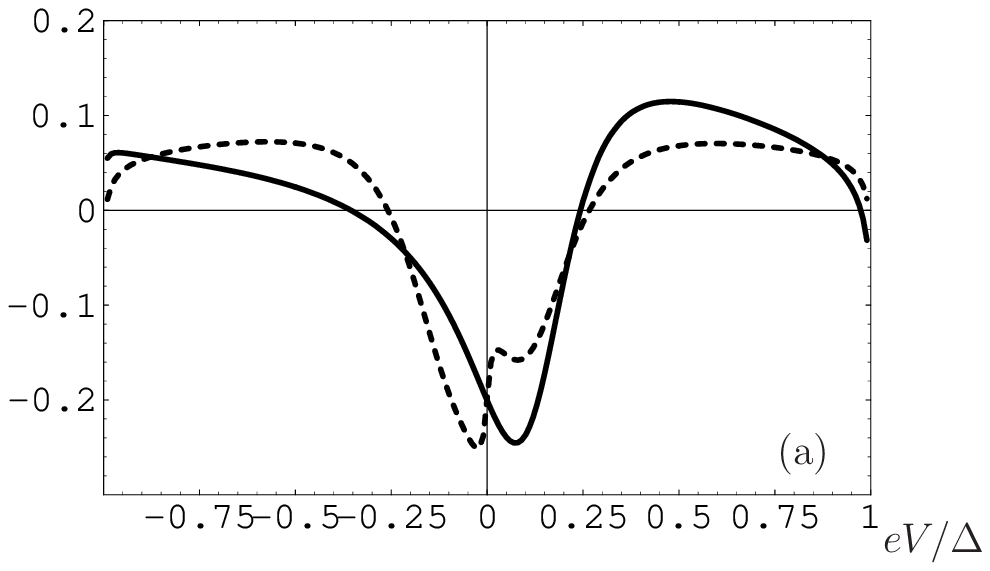}}
            \centerline{\includegraphics[clip=true,width=2.4in]{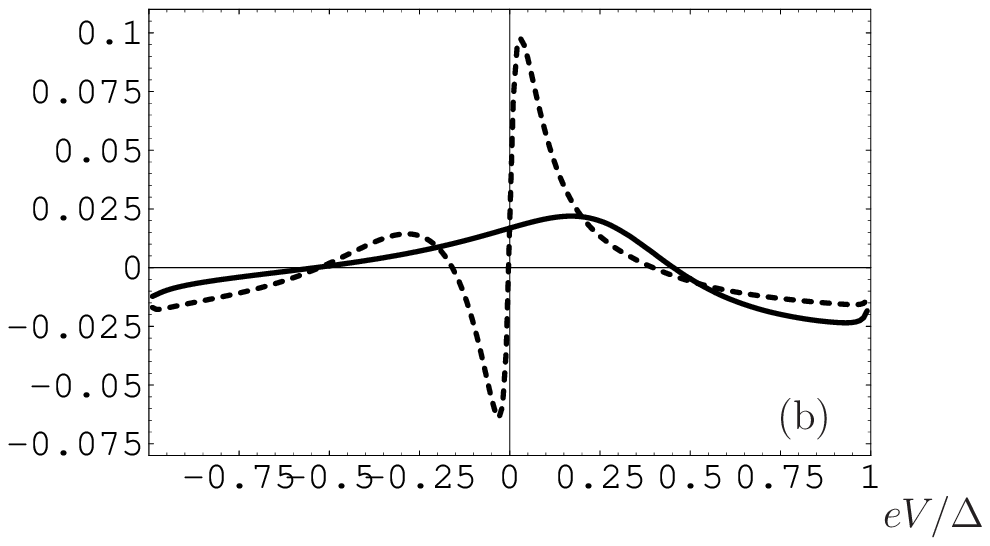}}
   \caption{Joint currents $j^c_s + j^c_{SR}$ (solid curves) and $j^c_s+j^c_{LR}$ (dashed curves)
as functions of $eV/\Delta$. For panel (a) $d_F=1$, $d_N=2$ and for panel (b) $d_F=2$, $d_N=1$.
The other parameters are the same as in Fig.~\ref{current_V}.}
\label{nonlinear}
\end{figure}

The possibility to extract singlet and triplet components of the proximity-induced anomalous Green's function
in the interlayer is not the only motivation to study the Josephson current under spin-dependent quasiparticle
distribution. Being an easily controllable parameter, voltage $V$ gives a possibility to obtain
highly nonlinear characteristics $j(V)$ with a number of $0$-$\pi$ transitions, which can be essential for
superconducting electronics. As it was already mentioned above, by choosing the appropriate orientation
of $P_l$ and $P_r$ magnetizations, one can, in principle, "turn off" either $j_{SR}$ or $j_{LR}$ current contribution.
When the full current through the junction is given by joint contribution of $j_{LR}$ and $j_s$ or $j_{SR}$ and $j_s$,
respectively. The corresponding full currents are demonstrated in Fig.~\ref{nonlinear}. Panel (a) reperesents the case
of short enough ferromagnetic layer $d_F=1$, while panel (b) corresponds to $d_F=2$. It is seen from panel (b) that
in this case the main contribution to the current is given by $j_{LR}$, at least for small enough voltages.
It is worth to note that it may be not easy to adjust $P_l$ and $P_r$ magnetizations in such a way that
only one of the components $j_{SR}$ and $j_{LR}$ flows. In fact, it is not necessary in order to obtain highly
nonlinear $j(V)$ characteristics. For this purpose it is enough to create any spin-dependent quasiparticle distribution
in the interlayer region. The internal structure of the anomalous Green's function $F_{LR}$ can also be studied
separately for long enough ferromagnetic interlayers.

\section{S/N/S junction with magnetic interfaces}

\label{SNS}

In this section Josephson current is studied for S/N/S junction with magnetic S/N interfaces
under the condition of spin-dependent quasiparticle distribution in the interlayer. The model is
already described in Sec.~\ref{model}. The anomalous Green's function in the interlayer is found
up to the first order in S/N conductance $\tilde G_T$
according to Eqs.~(\ref{usadel_lin}) and (\ref{boundary_lin}) (assuming $\bm h = 0$). In general,
the condensate penetrating into the interlayer region is comprised of two types of electron pairs: with opposite
electron spins and with parallel electron spins. However, due to the absence of ferromagnetic
elements in the interlayer region they have the same characteristic decay length.
The both types of pairs occur in the system if there is a reason for spin-flip there.
For example, it is the case if the magnetization vectors of the both interfaces
are not parallel $\bm m_l \nparallel \bm m_r$. If $\bm m_l || \bm m_r$, then only the pairs with opposite
electron spins, generated by the singlet superconductor, occur in the interlayer region. In order to make
the formulae less cumbersome we give final expressions only for the case $\bm m_l || \bm m_r \equiv \bm m$. Then,
at the left ($\alpha=+1$) and right ($\alpha=-1$) S/N interfaces the singlet part of the anomalous Green's function
takes the following form
\begin{eqnarray}
f_s^R=f_{s1}e^{-i\alpha \chi /2}+f_{s 2}e^{i\alpha \chi /2}
\enspace ,
~~~~~~~~~~~
\nonumber \\
\nonumber \\
f_{s1}=\frac{2i \pi G_T \sinh \Theta_S^R}{\sigma_N Z}\left[ \lambda_N+
\frac{1}{\lambda_N}\left( \frac{G_\phi}{\sigma_N} \right)^2 \right]  \sinh[2\lambda_N d_N]
\enspace ,
\nonumber \\
\nonumber \\
f_{s2}=\frac{4i \pi G_T \sinh \Theta_S^R}{\sigma_N Z}\left[ \lambda_N-
\frac{1}{\lambda_N}\left( \frac{G_\phi}{\sigma_N} \right)^2 \right]  \sinh[\lambda_N d_N]
\enspace ,
\nonumber \\
\nonumber \\
Z=4\lambda_N^2 \sinh^2[\lambda_N d_N] + 8 \left( \frac{G_\phi}{\sigma_N} \right)^2
\left( \cosh^2[\lambda_N d_N] +1 \right) +
\nonumber \\
4 \left( \frac{G_\phi}{\sigma_N} \right)^4 \frac{\sinh^2[\lambda_N d_N]}{\lambda_N^2 }
\label{fs_SNS}
\enspace ,
~~~~~~~~~~~~~~~~~~
\end{eqnarray}
where $\lambda_N$ is determined below Eq.~(\ref{fs_SNFNS}).

The triplet component of the anomalous Green's function has only $z$-component and takes the form
\begin{eqnarray}
\bm f_t^R=\left( 0,0,f_z \right)
\enspace ,~~~~~~~~~~~~~~~
\nonumber \\
\nonumber \\
f_z=f_{z1}e^{-i\alpha \chi/2}+f_{z2}e^{i\alpha \chi/2}
\enspace , ~~~~~~~~~
\nonumber \\
\nonumber \\
f_{z1}=\frac{4\pi G_\phi G_T \sinh \Theta_S^R}{\sigma_N^2 Z} \times ~~~~~~~~~~~~~~~
\nonumber \\
\left( \left[ 1+
\frac{1}{\lambda_N^2}\left( \frac{G_\phi}{\sigma_N} \right)^2 \right]\sinh^2[\lambda_N d_N]+2 \right)
\enspace ,
\nonumber \\
\nonumber\\
f_{z2}=\frac{8\pi G_\phi G_T \sinh \Theta_S^R}{\sigma_N^2 Z}\cosh[\lambda_N d_N]
\label{ft_SNS}
\enspace ,
\end{eqnarray}
where $Z$ is determined in Eq.~(\ref{fs_SNS}). Physically, $f_{s,z1}$ are generated by the proximity effect
at the same S/N interface and $f_{s,z2}$ are extended from the opposite S/N interface.

Just as in the previous section, in order to generate a spin-dependent quasiparticle distribution
in the interlayer, additional electrodes are attached to it. The principal scheme is the same as before except
for the fact that there is only one normal region in the considered system. Therefore, we assume that
the interlayer is attached to two additional normal electrodes $N_b$ and $N_t$ and
electrode $N_b$ has insertion $P$ made of a strongly ferromagnetic
material. The unit vector aligned with the magnetization of $P$ is denoted by $\bm M$.
Again, if voltage $2V$ is applied between the electrodes $N_b$ and $N_t$,
then the electric potentials for spin-up and spin-down electrons
in the $N_b$ region, inclosed between $P$ and the normal interlayer, counted from the level of
the superconducting leads are $V_\uparrow=(V_b-V_t)/2=V$ and $V_\downarrow=(V_t-V_b)/2=-V$.
The distribution functions for spin-up and spin-down electrons in this region are close
to the equilibrium form (with different electrochemical potentials). In a matrix form the distribution
function is expressed by Eq.~(\ref{distribution_normal}) with the substitution $\bm M$ for $\bm M_l$.

Now we can obtain the distribution function in the interlayer, which enters
the current [Eq.~(\ref{current_simp})]. Again, for simplicity we assume that $g_t \ll 1$. Consequently,
the dissipative current flowing through $N_b/N/N_t$ junction is negligible and, therefore, the $y$-dependence
of the distribution function in the interlayer region can be disregarded. For simplicity we assume below that
$\bm M || \bm m$. Under this condition the distribution function
$\hat \varphi^{(0)}$ in the interlayer calculated according to Eq.~(\ref{distribution_eq}) at $\bm h=0$
supplemented by boundary conditions at S/N interfaces (\ref{boundary_distrib_SN})
is spatially constant and equal to its value coming from $N_b$ region. If $\bm M \nparallel \bm m$, then
the spatially constant distribution function does not satisfy boundary conditions (\ref{boundary_distrib_SN})
any more. In this case the problem become two-dimensional and much more complicated.

Now we are able to calculate the Josephson current through the junction according to Eq.~(\ref{current_simp}).
After substitution of the expression for the singlet part of the anomalous Green's
function [Eq.~(\ref{fs_SNS})] and the scalar part of the distribution function [Eqs.~(\ref{distribution_normal})
and (\ref{symmetry})] into first two terms of Eq.~(\ref{current_simp}), the contribution of the SCDOS
singlet part takes the form
\begin{eqnarray}
j_s=\frac{2iG_T^2 \sin \chi}{e \sigma_N} \int \limits_{-\infty}^\infty \Delta^2 d \varepsilon
\tilde \varphi_0 (\varepsilon)\times~~~~~~~~~~~
\nonumber \\
\frac{\lambda_N \sinh [\lambda_N d_N](1-\left[ \frac{G_\phi}
{\sigma_N \lambda_N} \right]^2)}{
[(\varepsilon+i\delta)^2-\Delta^2]Z(\varepsilon)}
\label{js_SNS}
\enspace , ~~~~~~~
\end{eqnarray}
where $Z(\varepsilon)$ is determined in Eq.~(\ref{fs_SNS}).

The current flowing through the SCDOS triplet part and expressed by the third term in Eq.~(\ref{current_simp})
takes the form [in order to obtain this expression one should substitute Eqs.~(\ref{ft_SNS}), (\ref{distribution_normal})
and (\ref{symmetry}) into Eq.~(\ref{current_simp})]
\begin{eqnarray}
j_t=\frac{4G_T^2 G_\phi \sin \chi}{e \sigma_N^2} \int \limits_{-\infty}^\infty \Delta^2 d \varepsilon
\tilde \varphi_t (\varepsilon)\times
\nonumber \\
\frac{\cosh [\lambda_N d_N]}{[(\varepsilon+i\delta)^2-\Delta^2]Z(\varepsilon)}.
\label{jt_SNS}
\end{eqnarray}

As opposed to the problem of S/NFN/S junction considered in the previous section, it is seen from Eq.~(\ref{jt_SNS})
that $j_t$ values at the left and right S/N interfaces are equal to each other.
As for the case of S/NFN/S junction, the part of current (\ref{current_simp}) generated by the term
$\propto \cosh \Theta_S^R \left[ \varphi_0^{(0)}(\varepsilon)+\varphi_0^{(0)}(-\varepsilon) \right]/2$ vanishes
due to the fact that the scalar part $\varphi_0^{(0)}$ of the distribution function in the interlayer
[Eq.~(\ref{distribution_normal})] is an odd function of quasiparticle energy. Further, under the conditions
$|eV|<\Delta$ and $T \ll \Delta$ the last term, generated by
$\propto \alpha G_{MR} \cosh \Theta_S^R \bm m \left[ \bm \varphi^{(0)}(\varepsilon)+
\bm \varphi^{(0)}(-\varepsilon) \right]/2$ also vanishes because this expression is an odd function
of quasiparticle energy at $|\varepsilon|<\Delta$ and is absent elsewhere. Taking into account
that $\bm m_l || \bm m_r || \bm M$ one can obtain from Eq.~(\ref{boundary_distrib_SN}) that
$\partial_x \hat \varphi^{(1)}=0$ at the S/N interfaces.
Therefore, $\hat \varphi^{(1)}$ is approximately constant in the interlayer. Moreover, this constant
is to be equal to zero in order to satisfy the condition $j^l=j^r$. Therefore, the full Josephson
current flowing through the junction is given by the sum of singlet [Eq.~(\ref{js_SNS})] and triplet
[Eq.~(\ref{jt_SNS})] SCDOS contributions.

\begin{figure}[!tbh]
  \centerline{\includegraphics[clip=true,width=2.2in]{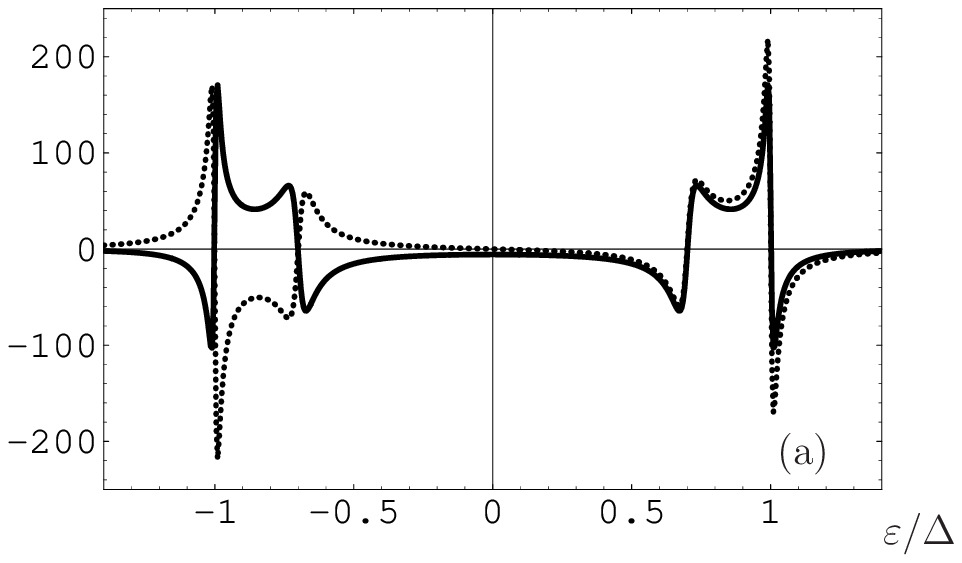}}
            \centerline{\includegraphics[clip=true,width=2.2in]{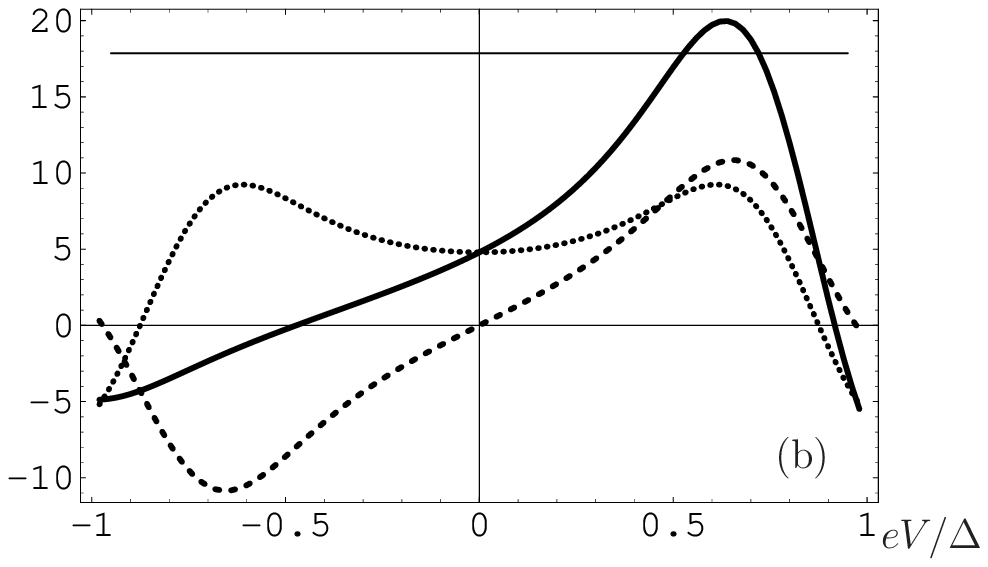}}
   \caption{(a) Functions $F_s$ (dotted line) and $F_t$ (solid line) as functions of $\varepsilon/\Delta$
for S/N/S junction with magnetic interfaces. (b) Full critical current (solid line) and its contributions
$j_s^c$ (dotted line) and $j_t^c$ (dashed line) as functions of $eV/\Delta$. For the both panels $d_N=0.5\xi_S$,
$G_\phi\xi_S/\sigma_N=0.35$ and $T=0.1\Delta$.}
\label{j+f_SNS}
\end{figure}

As for the previous case of S/NFN/S junction, $j^c_s$ is an even function of voltage $V$
applied to the additional electrodes and $j^c_t$ is an odd function of this voltage. Therefore,
contributions $j^c_s$ and $j^c_t$ can be extracted from an experimentally measurable Josephson current and, so, it
makes sense to discuss them separately. Panel (b) of Fig.~\ref{j+f_SNS} demonstrates the full
critical Josephson current and its contributions $j^c_s$ and $j^c_t$ as functions of $V$ for a typical set
of parameters (See caption to Fig.~\ref{j+f_SNS} for specific values). Functions $F_s(\varepsilon) \equiv
{\rm Im}\left[ f_{s2} (\varepsilon)/\sqrt{\Delta^2-(\varepsilon+i\delta)^2}\right]$
and $F_t(\varepsilon) \equiv {\rm Im}\left[ f_{z2} (\varepsilon)/
\sqrt{\Delta^2-(\varepsilon+i\delta)^2}\right]$ are represented in panel (a) of Fig.~\ref{j+f_SNS}
for the same set of parameters. As it is seen from the definitions given in Eqs.~(\ref{fs_SNS}) and (\ref{ft_SNS}),
these functions are proportional to the singlet and triplet components of the anomalous Green's function,
coming from the opposite S/N interface, and can be experimentally found by differentiating
the currents $j^c_s$ and $j^c_t$ with respect to voltage $V$, as it was
explained in the previous section.

\begin{figure}[!tbh]
   \begin{minipage}[b]{0.5\linewidth}
    \centerline{\includegraphics[clip=true,width=1.6in]{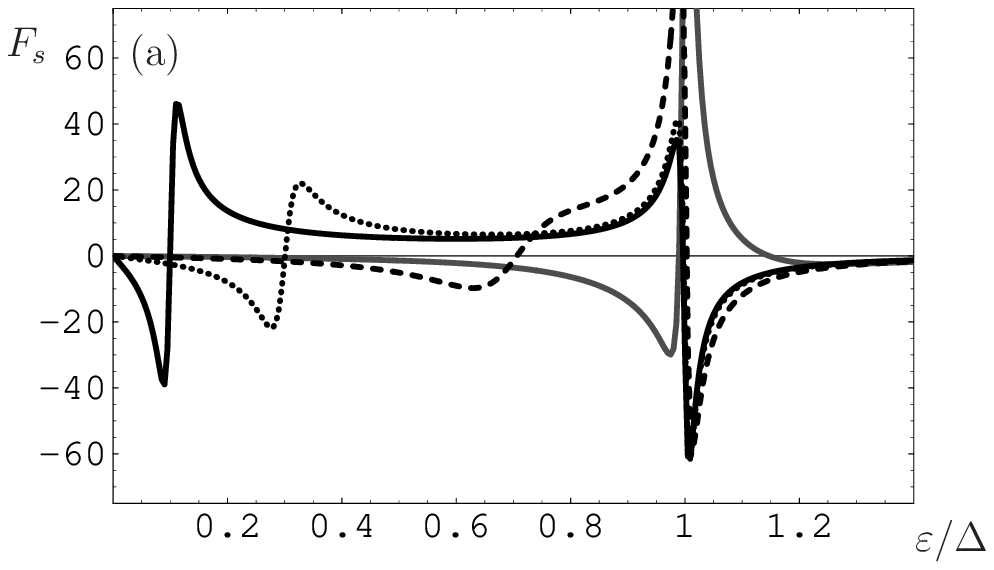}}
     \end{minipage}\hfill
\begin{minipage}[b]{0.5\linewidth}
    \centerline{\includegraphics[clip=true,width=1.6in]{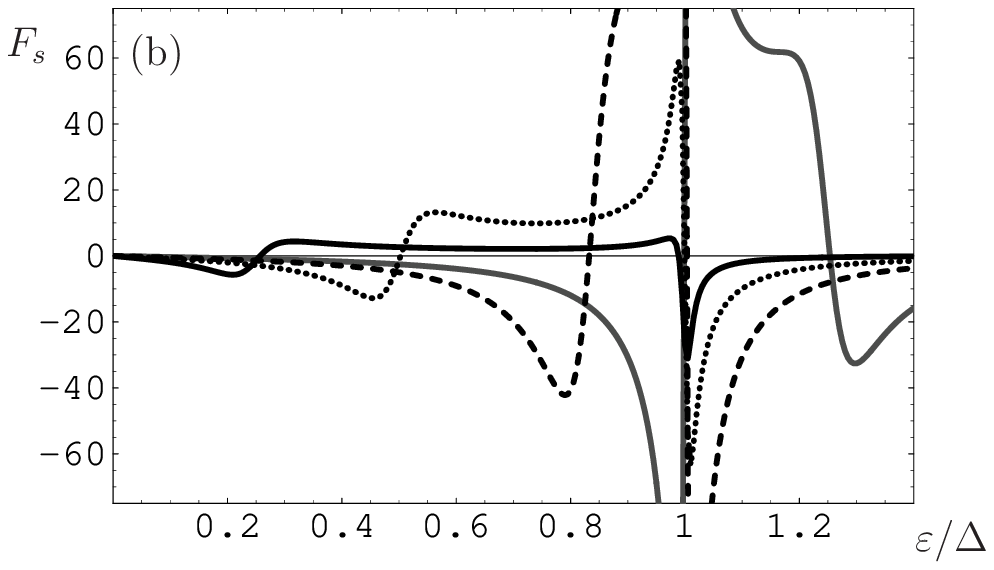}}
     \end{minipage}
    \begin{minipage}[b]{0.5\linewidth}
   \centerline{\includegraphics[clip=true,width=1.6in]{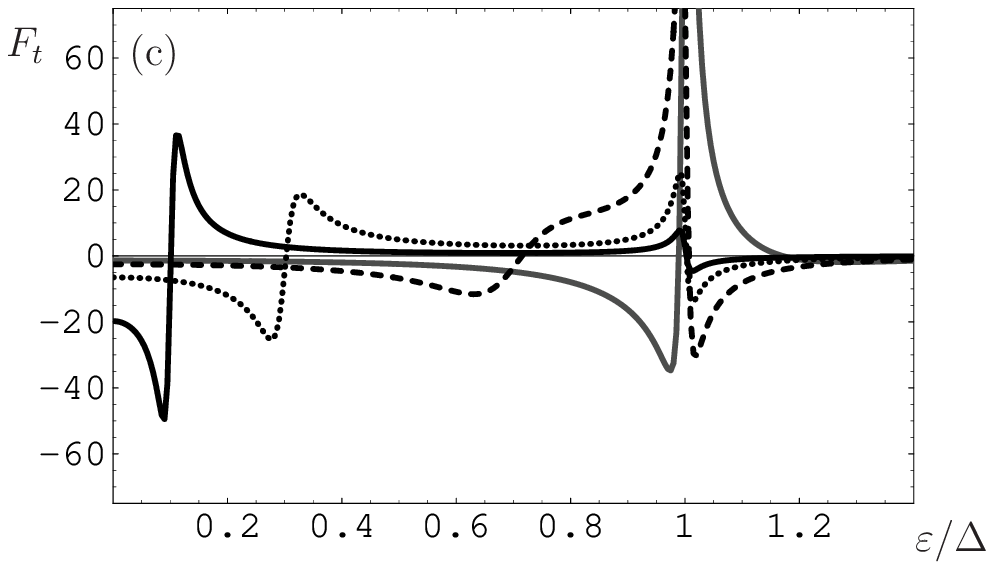}}
  \end{minipage}\hfill
\begin{minipage}[b]{0.5\linewidth}
    \centerline{\includegraphics[clip=true,width=1.6in]{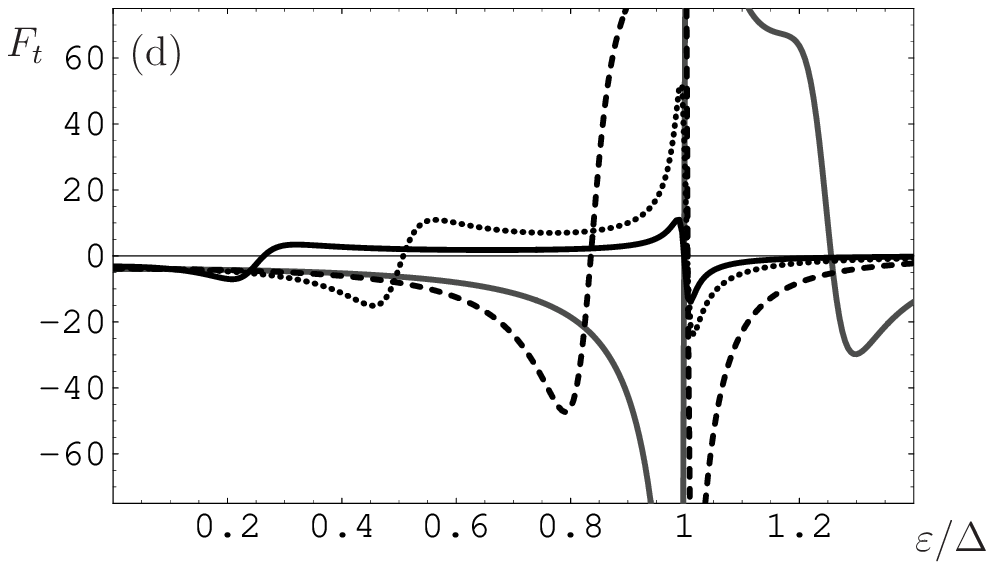}}
     \end{minipage}
   \caption{$F_s$ and $F_t$ as functions of $\varepsilon/\Delta$ for S/N/S junction with magnetic
interfaces. The upper row represents $F_s$, while the lower row demonstrates $F_t$. For panels (a) and (c) $d_N=\xi_S$
and different curves correspond to different values of $G_\phi \xi_S/\sigma_N=0.1$ (black solid curve),
0.3 (dotted curve), 0.7 (dashed curve) and 1.1 (gray solid curve). In panels (b) and (d) $G_\phi \xi_S/\sigma_N=0.5$
and different curves correspond to different $d_N/\xi_S=2$ (black solid curve), 1 (dotted curve), 0.6 (dashed curve)
and 0.4 (gray solid curve). $T=0.1\Delta$.}
\label{fs+ft_SNS}
\end{figure}

The characteristic shape of $F_s$ and $F_t$ dictates how $j_s$ and $j_t$ behave in dependence on $V$.
Upon discussing the characteristic features of $j_s$ and $j_t$ we only consider $V>0$ and, correspondingly,
$\varepsilon>0$ for $F_s$ and $F_t$.
The main characteristic features of $F_s$ and $F_t$, which are responsible for the current behavior,
are proximity induced dips at $\varepsilon_\phi \sim G_\phi \xi_S^2 \Delta / \sigma_N d_N$ (for the parameter
region $\varepsilon_\phi<\Delta$). These dips
are followed by abrupt changing of sign of the corresponding quantity. Fig.~\ref{fs+ft_SNS} shows $F_s$
and $F_t$ evolution with $G_\phi$ (left column) and with $d_N$ (right column). It is seen that upon
$G_\phi$ increasing the proximity induced dip shifts to higher energies. If the junction becomes shorter
the dip also shifts to the right and its integral height increases due to the fact that the proximity effect
is more pronounced for short junctions.

\begin{figure}[!tbh]
   \begin{minipage}[b]{0.5\linewidth}
    \centerline{\includegraphics[clip=true,width=1.6in]{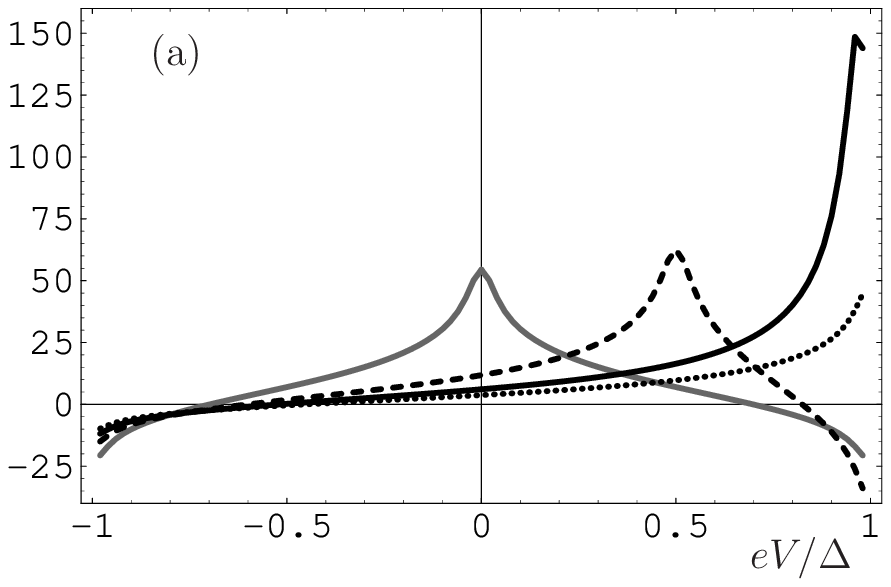}}
     \end{minipage}\hfill
\begin{minipage}[b]{0.5\linewidth}
    \centerline{\includegraphics[clip=true,width=1.6in]{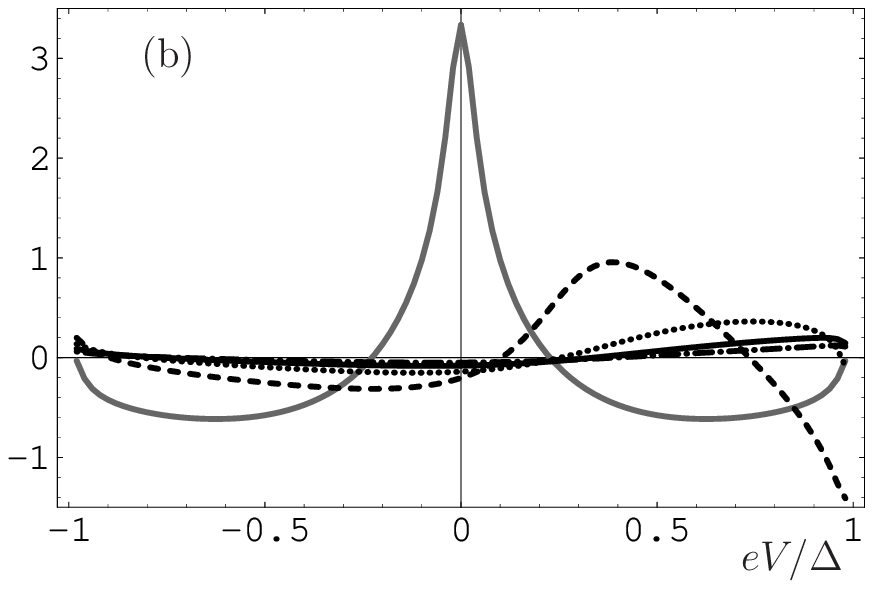}}
     \end{minipage}
    \begin{minipage}[b]{0.5\linewidth}
   \centerline{\includegraphics[clip=true,width=1.6in]{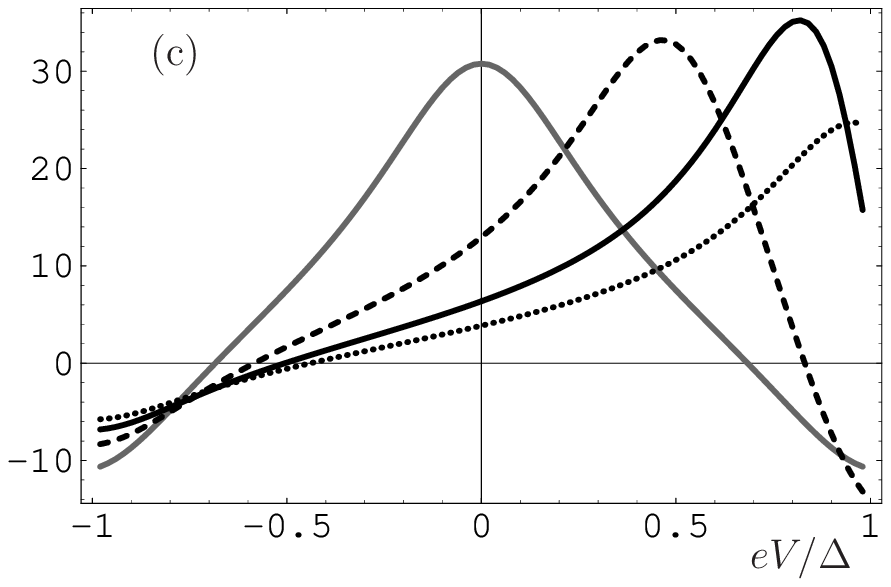}}
  \end{minipage}\hfill
\begin{minipage}[b]{0.5\linewidth}
    \centerline{\includegraphics[clip=true,width=1.7in]{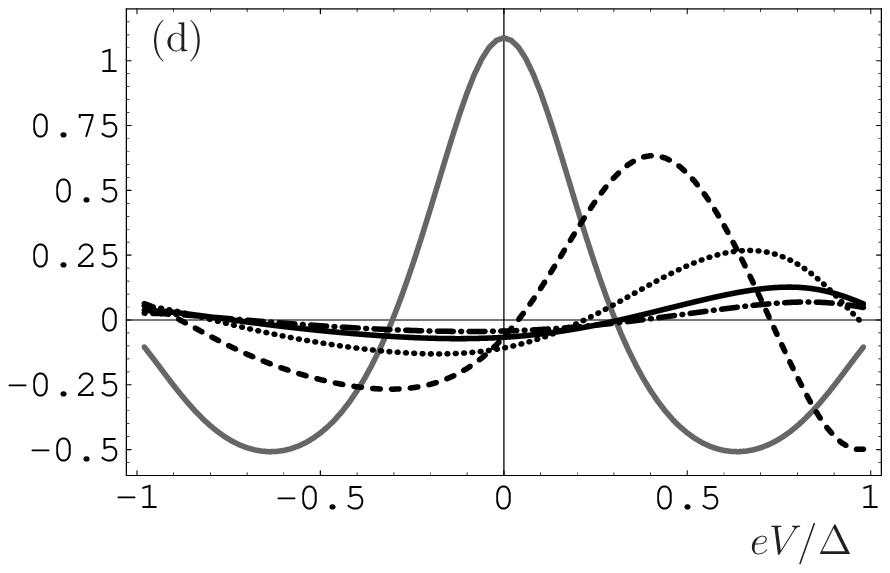}}
     \end{minipage}
   \caption{Full critical Josephson current for S/N/S junction with magnetic interfaces in dependence on $eV/\Delta$.
Panel (a) demonstrates the case of low-temperature junction, where the proximity effect is well-pronounced:
$T=0.01\Delta$, $d_N=0.3\xi_S$. $G_\phi\xi_S/\sigma_N=0.15$(dashed curve), $0.3$ (solid curve) and $0.45$ (dotted curve).
Panel (b) corresponds to longer junction at the same temperature: $T=0.01\Delta$,
$d_N=3\xi_S$. $G_\phi\xi_S/\sigma_N=1$(dashed curve), $2$ (dotted curve), $3$ (solid curve) and $4$ (dashed-dotted
curve). Panels (c) and (d) represent the same results as panels (a) and (b), respectively, but at
higher temperature $T=0.1\Delta$. For all the panels the gray solid line represents
$j_{nm}(V)$ for the corresponding set of the parameters.}
\label{jSNS_various}
\end{figure}

According to Eqs.~(\ref{current_simp}) and (\ref{distribution_normal}) at low enough temperatures only
the part of $F_s$, belonging to energy intervals $\left[ -\infty,-|eV| \right]$ and $\left[|eV|,+\infty \right]$,
contributes to $j^c_s$. Consequently, upon inreasing of $V$ the absolute value of $j^c_s$ grows
up to $V \sim \varepsilon_\phi$ and after that starts to decline
due to the sign changing of $F_s(\varepsilon)$ at $\varepsilon=\varepsilon_\phi$.
Analogously, only the part of $F_t$, belonging to energy interval
$\left[ -|eV|, |eV| \right]$, contributes to $j^c_t$. Therefore, the absolute value of $j^c_t$ also grows
up to $V \sim \varepsilon_\phi$ and declines after that. The described behavior is characteristic for
the absolute value of $j^c_s$ and $j^c_t$ as for $eV>0$, so as for $eV<0$. However, due to the fact that
$j^c_s$ is symmetric and $j^c_t$ is antisymmetric function of $V$, the total Josephson current
is highly nonsymmetric with respect to $V$, as it is seen in Figs.~\ref{j+f_SNS} and \ref{jSNS_various}.
While for $eV<0$ the contributions of $j^c_s$ and $j^c_t$ partially compensate each other leading to suppression
of the full current and $0-\pi$-transition at some finite $V$, they are added
for $eV>0$ resulting in the considerable current enhancement. The value of $eV$, where the peak in the critical
current is located, can be used for experimental estimate of spin-mixing parameter $G_\phi$, characterizing
the magnetic interface, because $eV_p \sim \varepsilon_\phi$. For short enough junctions with $d_N < \xi_S$
the current value can even exceed the critical current value for S/N/S junction with nonmagnetic S/N interfaces
($G_\phi=0$) and the same S/N interface conductance $G_T$ for some voltage range. It is worth to note
here that such an enhancement is only possible for finite $V$, when the triplet part of SCDOS $F_t$
contributes to the current. At $V=0$ the critical Josephson current through S/N/S junction with
magnetic interfaces $G_\phi \neq 0$ is always lower than the corresponding current fot S/N/S junction
with nonmagnetic intefaces but the same interface conductance $G_T$ (this statement is valid for the entire
range of parameters we consider).

Let us denote the value of the critical current for S/N/S junction
with nonmagnetic intefaces and interface conductance $G_T$ by $j_{nm}(V)$. In the framework of the microscopic
model of S/N interface considered in Appendix \ref{B}, comparison between $j(V)$ and $j_{nm}(V)$
physically corresponds to comparison between the Josephson currents in the system
with a thin magnetic layer between N and I and without it. The value
$j_{nm}(V=0)$ is shown in Fig.~\ref{j+f_SNS}(b) by the horizontal line. It is seen that a small
excess of the total current $j^c$ over $j_{nm}(V=0)$ takes place for some voltage range.
However, the excess can be much greater, the current
at finite $V$ can exceed the equilibrium current $j_{nm}(V=0)$ for nonmagnetic S/N/S junction
more than twice. Such a case is illustrated in Fig.~\ref{jSNS_various}(a). Maximal excess can be expected
for short junctions with $\varepsilon_\phi \approx \Delta$, where the proximity effect in $F_t$ is most
pronounced and the proximity induced dip at $\varepsilon_\phi$ merges with the coherence peak at
$\Delta$ thus greatly enhancing $F_t$ value in the subgap region. In addition the temperature should
be low enough in order to avoid temperature smearing of the effect.

For longer junctions with $d_N \gtrsim \xi_S$
the current does not exceed the equilibrium value $j_{nm}(V=0)$ because of weaker proximity effect in the interlayer
region, as it is illustrated in Fig.~\ref{jSNS_various}(b). For all the panels of Fig.~\ref{jSNS_various}
the gray solid line represents $j_{nm}(V)$ for the corresponding set of the parameters. It is worth to note here
that dependencies $j_{nm}(V)$ on $V$ are qualitatively very similar to the current discussed in
Ref.~\onlinecite{wilheim98} for nonmagnetic S/N/S junction under nonequilibrium quasiparticle distribution
in the normal interlayer. Indeed, at $G_\phi=0$ triplet part of SCDOS $F_t$ is absent and, consequently, the vector part
of the distribution function (\ref{distribution_normal}) does not contribute to the current. The singlet part of
this distribution function is formally equivalent to the nonequilibrium distribution \cite{wilheim98} for a narrow
normal interlayer (a wire or a constriction). Full quantitative agreement between our results for
$j_{nm}(V)$ and the results of Ref.~\onlinecite{wilheim98} cannot be reached because they are
obtained for somewhat different parameter ranges.

Panels (c) and (d) of Fig.~\ref{jSNS_various} show the results for the current
at higher temperature $T=0.1\Delta$. Panel (c) corresponds to shorter junction with $d_N=0.3\xi_S$, while panel (d)
demonstrates the case of longer junction with $d_N=3\xi_S$. It is seen that for short junction, where the effect
of current enhancement is well pronounced at low temperatures, raising of the temperature suppresses the effect.
The reason is that distribution function (\ref{distribution_normal}) smears upon raising of the temperature. Consequently,
not only the part of $F_t$ corresponding to $|\varepsilon|<\varepsilon_\phi$, but also
some region of higher energies, where $F_t$ has opposite sign, is involved into the current $j_t$ now.
This leads to partial compensation of $F_t$ parts with different signs.

\begin{figure}[!tbh]
  \centerline{\includegraphics[clip=true,width=2.2in]{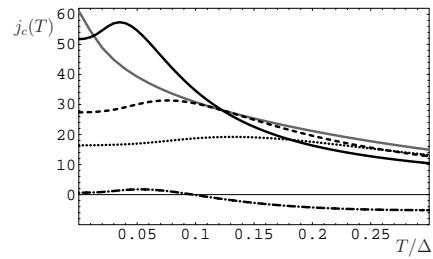}}
   \caption{Full critical Josephson current for S/N/S junction with magnetic interfaces in dependence on temperature
for several different voltages $V$. $d_N=0.3\xi_S$, black solid curve: $G_\phi\xi_S/\sigma_N=0.3$, $eV/\Delta=0.85$;
dashed curve: $G_\phi\xi_S/\sigma_N=0.3$, $eV/\Delta=0.7$; dotted curve: $G_\phi\xi_S/\sigma_N=0.3$, $eV/\Delta=0.5$;
dashed-dotted curve: $G_\phi\xi_S/\sigma_N=0.15$, $eV/\Delta=0.83$. Gray solid curve represents
the temperature dependence of $j_{nm}(V=0)$ for the corresponding set of parameters.}
\label{jSNS_T}
\end{figure}

Although the maximal value of the critical Josephson current, which can be reached at a finite $V$, is suppressed by
temperature, the current dependence on $T$ at a particular voltage $V$ can be quite interesting. Fig.~\ref{jSNS_T}
demonstrates how the current depends on temperature at several specified values of voltage $V$ for the case of short
junction with $d_N=0.3\xi_S$. The gray solid line represents the dependence of $j_{nm}(V=0)$ on temperature
and is given for comparison of our results with the equilibrium nonmagnetic case. It is well known, that
the Josephson current for equilibrium nonmagnetic S/N/S junction declines upon raising of temperature rather sharply,
as it is demonstrated by the gray solid curve. On the contrary, the current at finite $V$ for a S/N/S with magnetic
interfaces can even grow up to some temperature and only after that start to decline. The qualitative explanation
of this fact is the following. The main
contribution to the Josephson current in nonmagnetic equilibrium S/N/S junction is given by high peak of SCDOS
located at low energies. Consequently, the temperature smearing of the equilibrium distribution function
$\tanh \varepsilon/2T$ crucially reduces the current. At the same time, the main contribution to $j_t$ is given by
the energies up to $\varepsilon \sim |eV|$ and under the condition that $|eV|<\varepsilon_\phi$ the temperature
smearing of the distribution function involves higher energies, where the absolute value of $F_t$ even larger,
in the current transfer. In addition, at some voltage ranges the junction can manifest $0-\pi$ transition
in dependence on temperature.

\section{summary}

\label{summary}

In conclusion, we have theoretically investigated the Josephson current in weak links,
containing ferromagnetic elements, under the condition that the quasiparticle distribution in the weak link region
is spin-dependent. Two types of weak link are considered. The first system is a S/N/F/N/S junction with
complex interlayer composed of two normal metal regions and a middle layer made of a spiral ferromagnet,
sandwiched between them. The second considered system is a S/N/S junction with magnetic S/N interfaces.
In both cases spin-dependent quasiparticle distribution in the interlayer region is proposed to be created
by attachment of additional electrodes with ferromagnetic elements to the interlayer region and applying a voltage
$V$ between them. Interplay of the triplet superconducting correlations, induced in the interlayer by the proximity
with the superconducting leads, and spin-dependent quasiparticle distribution results in the appearence
of the additional contribution to the Josephson current $j_t$, carried by the triplet part of SCDOS.

It is shown that $j_t$ is an odd function of $V$, while the standard contribution $j_s$, carried by
the singlet part of SCDOS, is an even function. So, $j_t$ can be extracted from the full Josephson current
measured as a function of $V$. Further, it is demonstrated that derivative $dj_t/dV$ can provide direct information
about the anomalous Green's function describing the superconducting triplet correlations induced in the interlayer.
We show that in the S/N/F/N/S junction the contributions given by the short-range (SRTC) and long-range
(LRTC) components of triplet superconducting correlations in the interlayer can be measured separately.

For S/N/S junction with magnetic interfaces it is also obtained that the critical Josephson current at some finite $V$
can considerably exceed the current flowing through the equilibrium nonmagnetic S/N/S junction with the same
S/N interface transparency. This enhancement is due to the fact that the triplet component of SCDOS "works"
under spin-dependent quasiparticle distribution giving the additional contribution to the current, while it
does not take part in the current transfer for spin-independent quasiparticle distribution. In addition,
we have studied temperature dependence of the critical current in S/N/S junction with magnetic S/N interfaces.
As opposed to the case of equilibrium nonmagnetic S/N/S junction, where the current is monotonouosly suppressed
by temperature, in the considered case at a finite voltage $V$ it can at first rise with temperature and only
then start to decline.

The dependence of the full critical current on $V$ is typically highly nonlinear and strongly nonsymmetric
with respect to $V=0$ due to the interplay of $j_s$ and $j_t$. This also leads to appearence of a number of
$0$-$\pi$ transitions in the system upon varying controlling voltage $V$.

\begin{acknowledgments}
The authors acknowledge the support by RFBR Grant 09-02-00799-a.
\end{acknowledgments}

\appendix

\section{Microscopic calculation of anomalous Green's function and Josephson current in S/NFN/S junction}

\label{A}

In this Appendix we calculate the anomalous Green's functions $f_s$, $f_{SR}$ and $f_{LR}$
and the corresponding current contributions $j_s$, $j_{SR}$ and $j_{LR}$ in the framework of the most simple microscopic
model for N/F/N interlayer. We assume the NF interfaces to be absolutely transparent. This approximation simplifies
the calculations significantly, but does not influence qualitatively our main conclusions.
For this case the boundary conditions at $x=\mp d_F/2$ take the form
\begin{eqnarray}
\check g_N = \check g_F \enspace ,~~~~~~
\nonumber \\
\sigma_N \partial_x \check g_N = \sigma_F \partial_x \check g_F
\label {boundary_FN}
\enspace .
\end{eqnarray}
As far as we only need anomalous Green's functions to the first order in S/N interface transparency, the above
boundary conditions should be linearized. Then for retarded and advanced Green's functions they read as follows
\begin{eqnarray}
\hat f_N^{R,A} = \hat f_F^{R,A} \enspace ,~~~~~~
\nonumber \\
\sigma_N \partial_x \hat f_N^{R,A} = \sigma_F \partial_x \hat f_F^{R,A}
\label{FN_lin}
\enspace .
\end{eqnarray}

The boundary conditions for the distribution function at N/F interface to the considered accuracy take the form
\begin{eqnarray}
\hat \varphi_F = \hat \varphi_N
\enspace ,~~~~~~~~
\nonumber \\
\sigma_F \partial_x \hat \varphi_F = \sigma_N \partial_x \hat \varphi_N
\label{boundary_distrib_FN}
\enspace .
\end{eqnarray}

The singlet part of the anomalous Green's function, calculated according to Eqs.~(\ref{usadel_lin}), (\ref{boundary_lin})
and (\ref{FN_lin}), at the left ($\alpha=+1$) and the right ($\alpha=-1$) S/N interfaces takes the following form
\begin{eqnarray}
f_s^R=\frac{i\pi G_T}{\sigma_N \lambda_N}\tanh \phi_N \sinh \Theta_S^R e^{-i\alpha \chi /2}+
\frac{i\pi G_T \sinh \Theta_S^R }{2\sigma_F \cosh^2 \phi_N}\times~~~~
\nonumber \\
\left[ \frac{\cos (\chi /2) \cosh \phi_+}
{\lambda_+ \sinh \phi_+ + \rho \cosh \phi_+}-\frac{i \alpha \sin (\chi /2) \sinh \phi_+}
{\lambda_+ \cosh \phi_+ + \rho \sinh \phi_+}+ \right.~~~~~~~
\nonumber \\
\left. \frac{\cos (\chi /2) \cosh \phi_-}
{\lambda_- \sinh \phi_- + \rho \cosh \phi_-}-\frac{i \alpha \sin (\chi /2) \sinh \phi_-}
{\lambda_- \cosh \phi_- + \rho \sinh \phi_-}  \right],~~~~~~~
\label{fs_SNFNS}
\end{eqnarray}
where $\lambda_\pm=\sqrt{h/D}(1\mp i)$, $\lambda_N=\sqrt{-2i(\varepsilon+i\delta)/D}$,
$\phi_\pm=\lambda_\pm d_F/2$, $\phi_N=\lambda_N d_N/2$ and $\rho=(\sigma_N/\sigma_F)\lambda_N \tanh \phi_N$.

The results for triplet components $f_{SR}$ and $f_{LR}$ are the following
\begin{widetext}
\begin{eqnarray}
f_{SR}= -\frac{i\pi G_T \sinh \Theta_S^R }{2\sigma_F \cosh^2 \phi_N}
\left[ \frac{\cos (\chi /2) \cosh \phi_+}
{\lambda_+ \sinh \phi_+ + \rho \cosh \phi_+}-\frac{i \alpha \sin (\chi /2) \sinh \phi_+}
{\lambda_+ \cosh \phi_+ + \rho \sinh \phi_+}-\right.~~~~~~~~~~~~~~~~
\nonumber \\
\left.\frac{\cos (\chi /2) \cosh \phi_-}
{\lambda_- \sinh \phi_- + \rho \cosh \phi_-}+\frac{i \alpha \sin (\chi /2) \sinh \phi_-}
{\lambda_- \cosh \phi_- + \rho \sinh \phi_-}  \right] \enspace ,~~~~~~~~~~~~~~~
\nonumber \\
f_{LR}= -\frac{i\pi G_T \sinh \Theta_S^R }{2\sigma_F \cosh^2 \phi_N}\left\{ \frac{\Theta' i\sin (\chi /2)
\cosh \phi_t}{\rho \cosh \phi_t + \lambda_t \sinh \phi_t}\left[ \frac{\sinh \phi_+}{\lambda_+ \cosh \phi_+ +
\rho \sinh \phi_+}- \frac{\sinh \phi_-}{\lambda_- \cosh \phi_- +
\rho \sinh \phi_-} \right] - \right.
\nonumber \\
\left. \frac{\alpha \Theta' \cos(\chi /2)
\sinh \phi_t}{\rho \sinh \phi_t + \lambda_t \cosh \phi_t}\left[ \frac{\cosh \phi_+}{\lambda_+ \sinh \phi_+ +
\rho \cosh \phi_+}- \frac{\cosh \phi_-}{\lambda_- \sinh \phi_- +
\rho \cosh \phi_-} \right] \right\}
\label{fSR_fLR}
\enspace ,
\end{eqnarray}
\end{widetext}
where $\lambda_t=\sqrt{{\Theta'}^2-2i(\varepsilon+i\delta)/D}$ and $\phi_t=\lambda_t d_F /2$.

The fact that $f_{SR}$ rapidly decays in the ferromagnetic region and, consequently, represents
the SRTC can be easily seen from Eq.~(\ref{fSR_fLR}) in the limit of thick enough F layer: $d_F/\xi_F \gg 1$.
To the leading order in the parameter $e^{-d_F/\xi_F}$ for quantities $f_{SR}^l$ and $f_{SR}^r$,
defined by Eq.~(\ref{fSR_divide}), one obtains from Eq.~(\ref{fSR_fLR})
\begin{eqnarray}
f_{SR}^l=-\frac{i\pi G_T \sinh \Theta_S^R }{2\sigma_F \cosh^2 \phi_N}
\left[ \frac{1}{\lambda_+ + \rho}-\frac{1}{\lambda_- + \rho} \right],
~~~~~
\nonumber \\
f_{SR}^r=-\frac{i\pi G_T \sinh \Theta_S^R }{\sigma_F \cosh^2 \phi_N}
\left[ \frac{\lambda_+ e^{-\lambda_+ d_F}}{(\lambda_+ + \rho)^2}-
\frac{\lambda_- e^{-\lambda_- d_F}}{(\lambda_- + \rho)^2} \right].
\label{fSR_thick}
~~~
\end{eqnarray}

At the same regime the corresponding components of $f_{LR}$ take the following form
\begin{eqnarray}
f_{LR}^l=\frac{i\pi G_T \sinh \Theta_S^R }{2\sigma_F \cosh^2 \phi_N}
\left[ \frac{1}{\lambda_+ + \rho}-\frac{1}{\lambda_- + \rho} \right]\times~~~~~
\nonumber \\
\frac{\Theta'(\lambda_t \cosh [2\phi_t]+\rho \sinh [2\phi_t])}
{(\rho \cosh \phi_t + \lambda_t \sinh \phi_t)(\rho \sinh \phi_t +
\lambda_t \cosh \phi_t)}
\enspace ,~~~~~
\nonumber \\
f_{LR}^r=-\frac{i\pi G_T \sinh \Theta_S^R }{2\sigma_F \cosh^2 \phi_N}
\left[ \frac{1}{\lambda_+ + \rho}-\frac{1}{\lambda_- + \rho} \right]\times~~~~~
\nonumber \\
\frac{\Theta'\lambda_t}
{(\rho \cosh \phi_t + \lambda_t \sinh \phi_t)(\rho \sinh \phi_t +
\lambda_t \cosh \phi_t)}
\label{fLR_thick}
\enspace .~~~
\end{eqnarray}
As it is seen, $f_{LR}^r$ does not contain the small factor $e^{-d_F/\xi_F}$ in the leading
approximation and, therefore, $f_{LR}$ describes the LRTC.
The characteristic decay length of $f_{LR}$
in the F layer is $|\lambda_t|^{-1}$.

To the considered accuracy the singlet component of the anomalous Green's function also
decays at the distance $\sim \xi_F$ in the F layer, just as the SRTC $f_{SR}$ does.
In the regime $d_F/\xi_F \gg 1$ it can be obtained from Eq.~(\ref{fs_SNFNS}) that
\begin{eqnarray}
f_s^l=\frac{i\pi G_T}{\sigma_N \lambda_N}\tanh \phi_N \sinh \Theta_S^R+~~~~~~~~~~
\nonumber \\
\frac{i\pi G_T \sinh \Theta_S^R }{2\sigma_F \cosh^2 \phi_N}
\left[ \frac{1}{\lambda_+ + \rho}+\frac{1}{\lambda_- + \rho} \right]
\enspace ,~~~~~
\nonumber \\
f_s^r=\frac{i\pi G_T \sinh \Theta_S^R }{\sigma_F \cosh^2 \phi_N}
\left[ \frac{\lambda_+ e^{-\lambda_+ d_F}}{(\lambda_+ + \rho)^2}+
\frac{\lambda_- e^{-\lambda_- d_F}}{(\lambda_- + \rho)^2} \right].
\label{fs_thick}
\end{eqnarray}

Substituting Eqs.~(\ref{fs_SNFNS}) and (\ref{fSR_fLR}) together with the expressions for
the vector part of the distribution function [Eqs.~(\ref{distribution_normal}) and (\ref{symmetry})]
into Eq.~(\ref{current_simp}) one can find
\begin{eqnarray}
j_s=\frac{G_T^2 \sin \chi}{8 e \sigma_F} \int \limits_{-\infty}^\infty \frac{i \Delta^2 d \varepsilon
\tilde \varphi_0 (\varepsilon)}{\left[(\varepsilon+i\delta)^2-\Delta^2 \right]\cosh^2 \phi_N}\times
\nonumber \\
\left[ \frac{1}{\lambda_+ \tanh \phi_+ + \rho_N }-
\frac{\tanh \phi_+}{\lambda_+ + \rho_N \tanh \phi_+}+\right.~~~~~
\nonumber \\
\left.\frac{1}{\lambda_- \tanh \phi_- + \rho_N }-
\frac{\tanh \phi_-}{\lambda_-  + \rho_N \tanh \phi_-} \right],~~~~~~
\label{js_SNFNS}
\end{eqnarray}

\begin{eqnarray}
j_{SR}=-\frac{G_T^2 \sin \chi}{8 e \sigma_F} \int \limits_{-\infty}^\infty \frac{i \Delta^2 d \varepsilon
\tilde \varphi_t (\varepsilon)}{\left[\Delta^2-(\varepsilon+i\delta)^2 \right]\cosh^2 \phi_N}\times
\nonumber \\
\left[ \frac{1}{\lambda_+ \tanh \phi_+ + \rho_N }-
\frac{\tanh \phi_+}{\lambda_+ + \rho_N \tanh \phi_+}-\right.~~~~~
\nonumber \\
\left.\frac{1}{\lambda_- \tanh \phi_- + \rho_N }+
\frac{\tanh \phi_-}{\lambda_- + \rho_N \tanh \phi_-} \right],~~~~~~~
\label{jSR}
\end{eqnarray}
\begin{eqnarray}
j_{LR}=-\frac{G_T^2 \sin \chi}{8 e \sigma_F} \int \limits_{-\infty}^\infty \frac{i \Delta^2 d \varepsilon
\tilde \varphi_t (\varepsilon)}{\left[\Delta^2-(\varepsilon+i\delta)^2 \right]\cosh^2 \phi_N}\times
\nonumber \\
\left\{ \frac{\Theta'}{\rho + \lambda_t \tanh \phi_t}
\left[ \frac{\tanh \phi_+}{\lambda_+ +
\rho \tanh \phi_+}- \right. \right.
\nonumber \\
\left. \frac{\tanh \phi_-}{\lambda_- +
\rho \tanh \phi_-} \right] -\frac{\Theta'\tanh \phi_t}{\rho \tanh \phi_t + \lambda_t } \times
\nonumber \\
\left. \left[ \frac{1}{\lambda_+ \tanh \phi_+ +
\rho }- \frac{1}{\lambda_- \tanh \phi_- +
\rho } \right] \right\}.~~~~~
\label{jLR}
\end{eqnarray}

\section{Microscopic model of magnetic S/N interface}

\label{B}

Let us introduce the electronic scattering matrix $S^e$ associated to electrons
with spin $\sigma$ of the ${\rm n}^{{\rm th}}$ transmission channel. We assume that the
interface do not rotate an electron spin, that is the scattering matrix is diagonal in spin space.
\begin{equation}
S^e_{n\sigma}=\left(
\begin{array}{cc}
r^l_{n\sigma} & t^r_{n\sigma} \\
t^l_{n\sigma} & r^r_{n\sigma} \\
\end{array}
\right)
\label{Smatrix}
\enspace ,
\end{equation}
where $r^{l(r)}_{n\sigma}$ denotes the reflection amplitude at the left (right) side of the interface and
$t^{l(r)}_{n\sigma}$ the transmission amplitude from the left (right) side to the right (left) side
of the interface. Taking into account the constraints on $S^e$ resulting from the unitarity condition
$S^e S^{e\dagger} = 1$ and time reversal symmetry one can show that without any loss of generality
$S^e$ is entirely determined by the following parameters: the transmission probability $T_n$,
the degree of spin polarization $P_n$ and the spin-mixing angle $d \varphi_n^{l(r)}$. These parameters are defined
as $T_{n\sigma}=|t_{n\sigma}|^2=T_n(1+\sigma P_n)$ and ${\rm arg} [r^{l(r)}_{n\sigma}]=\varphi^{l(r)}_n+\sigma
(d \varphi^{l(r)}_n /2)$.

These parameters can be straightforwardly calculated in the framework of a microscopic model describing the interface.
Here we model S/N interface by an unsulating barrier I (with a transparency $T \ll 1$)
and a thin layer of a ferromagnetic metal, which is located between I and the normal interlayer. This layer
is supposed to provide a required value for the spin-mixing angle. Given the exchange field in the ferromagnetic
layer is small with respect to the Fermi energy $h \ll \varepsilon_F$, in the framework of this microscopic
model $T_n=T$, $P_n \approx 0$ and $d \varphi_n \approx 2 w_F h / v_F$, where $w_F$ is the length of the ferromagnetic
layer and $v_F$ is the corresponding Fermi velocity.

The main parameters entering magnetic boundary conditions (\ref{boundary_magnetic})
are connected to the microscopic parameters $T_n$, $P_n$ and $d \varphi_n^{l(r)}$ by the following
way \cite{cottet09}
\begin{equation}
G_T S=2G_q \sum \limits_n T_n
\label{G_T}
\enspace ,
\end{equation}
\begin{equation}
G_{MR} S=G_q \sum \limits_n T_n P_n
\label{G_MR}
\enspace ,
\end{equation}
\begin{equation}
G_\phi S=2G_q \sum \limits_n (T_n-1)d \varphi_n
\label{G_phi}
\enspace ,
\end{equation}
where $S$ is the junction area and $G_q=e^2/h$ is the quantum conductance. It is worth to note here
that boundary conditions (\ref{boundary_magnetic}) are the expansion in small $T_n$, $P_n$ and $d\varphi_n$
of the more general boundary conditions \cite{cottet09} and, consequently, Eq.~(\ref{boundary_magnetic})
is only valid if all these parameters are considerably less than unity. However, because of summation
over large number of transmisson channels, it does not mean that the parameters $G_T$, $G_{MR}$
and $G_\phi$ must be small. Let us estimate the value of dimensionless $\tilde G_\phi=G_\phi \xi_S /\sigma_N$,
which can be obtained in the framework of our microscopic model. For $T \ll 1$
\begin{equation}
\tilde G_\phi \approx -\frac{N \xi_S G_q}{S \sigma_N} d \varphi \sim -\frac{\xi_S}{l} d \varphi
\label{estimate_Gphi}
\enspace ,
\end{equation}
where $N$ is the number of transmission channels and $l$ is the mean free path.
$d \varphi$ means the average value of the spin-mixing
angle $d \varphi_n$. For rough estimates it is possible to take $d\varphi \approx 2w_F h /v_F$.

Our main results for S/N/S junction with magnetic interfaces are calculated for $\tilde G_\phi \sim 1$.
From Eq.~(\ref{estimate_Gphi}) it is seen that it is quite reasonable to expect such values of $\tilde G_\phi$
for a magnetic interface composed of an unsulating barrier and a weak ferromagnetic alloy with $w_F \ll \xi_F$.


\end{document}